\documentclass[12pt,a4paper,onecolumn]{article}
\usepackage{graphicx}
\usepackage{amssymb}
\begin{document}
\input epsf
\epsfverbosetrue
\setcounter{page}{1}
\pagenumbering{arabic}
\vspace{66pt}
\begin{center}
{\bf 
Multifragmentation {\it vs.} Evaporation {\it vs.} Binary-Decay \\
in Fragment Production 
}\\
\vspace{33pt}
{\bf
S. G. Mashnik$^{1}$,
K. K. Gudima$^{2}$,
M. I. Baznat$^{2}$,
}

\vspace{-1mm}
$^1$X-3, Los Alamos National Laboratory, Los Alamos, New Mexico 87545, USA\\
$^{2}$Institute of Applied Physics, Academy of Science of
Moldova, Chi\c{s}in\u{a}u, Moldova

\end{center}
\begin{center}
{\bf Abstract}
\end{center}
This paper presents part of an internal LANL Progress Report
on completion of the ``S" and ``G" versions of 
the improved Cascade-Exciton Model (CEM03.01) 
and the Los Alamos Quark-Gluon String Model (LAQGSM.03.01)
codes.
The ``S" versions consider fragmentation of compound nuclei produced 
after the preequilibrium stage of reactions for excitation energies above
$2\times A$ MeV using the Statistical Multifragmentation Model (SMM)
by Botvina {\it et al.} (``S" stands for SMM), while the ``G" versions
describe evaporation/fission stages of reactions using the
fission-like binary-decay model GEMINI of Charity {\it et al.}
(``G" stands for GEMINI) instead of using the 
the Generalized Evaporation Model GEM2 of Furihata
incorporated into the standard versions of these codes.
We present here an analysis of the recent
660 MeV p + $^{129}$I and 3.65 GeV p + $^{112}$Sn JINR  measurements,
of the new COSY data on 1.2 GeV p + (13 nuclei from Al to Th),
of the 300 MeV and 1 GeV p + $^{56}$Fe data measured at GSI in
inverse kinematics, and of the new GSI data on 
1 GeV/nucleon $^{124}$Xe and $^{136}$Xe + Pb.
To better understand the mechanisms of fragment
production, we discuss several calculated but not-yet-measured
kinematic characteristics of products of these reactions, which are 
predicted to be quite different by SMM, GEMINI, and GEM2. 
We find these kinematic quantities to be potentially useful in
differentiating these reaction mechanisms if they can be
measured in future experiments.
 
{\bf 1. Introduction}

For Proton Radiography (PRAD) 
as a radiographic probe for the Advanced Hydro-test Facility 
and other LANL applications, we have developed recently
(see, {\it e.g.}, \cite{GSI03}--\cite{PhotoCEM03} and references therein)
improved versions of the Cascade-Exciton Model (CEM) \cite{CEM}
and of the Los Alamos Quark-Gluon String Model (LAQGSM) \cite{LAQGSM}
codes as event generators to be used in MCNP6, MARS, and MCNPX
transport codes.

The latest versions of the event generators,
CEM03.01 and LAQGSM03.01, 
\cite{ResNote05}--\cite{CEM03.01}
have significantly improved IntraNuclear Cascade (INC) models,
updated preequilibrium, Fermi break-up,  and coalescence
models able to describe better than
their predecessors emission of complex particles and light fragments, 
and were extended to describe
photonuclear reactions up to tens of GeV.
On the whole, CEM03.01 and LAQGSM03.01 describe nuclear reactions
much better than their predecessors and other similar codes
available to the nuclear physics community.
They have been benchmarked on a variety of particle-particle,
particle-nucleus, and nucleus-nucleus reactions at energies from
10 MeV to 800 GeV per nucleon, and have
have been or are being incorporated as event generators into 
the transport codes MCNP6, MARS, and MCNPX.

However, both CEM03.01 and LAQGSM03.01 still have 
some problems in providing an accurate description of light- and 
medium-mass fragments produced from some nuclear reactions
on intermediate-mass targets, which cannot fission or fragment into
many channels using the GEM2 model. We address
this problem in two different ways \cite{ResNote06}:

1) By implementing into CEM03.01 and LAQGSM03.01 the
Statistical Multifragmentation Model (SMM)
by Botvina {\it et al.} 
\cite{SMM}--\cite{SMM_history},
to consider multifragmentation as a mode competitive
to evaporation of particles and light fragments, when the
excitation energy $U$ of a compound nucleus produced after the
preequilibrium stage of a reaction is above $2\times A$ MeV. 
This way, we have produced the ``S" version of our codes 
(``S" stands for SMM), CEM03.S1 and LAQGSM03.S1.

2) By replacing the Generalized Evaporation Model GEM2 of Furihata 
\cite{GEM2}--\cite{Furihata3}
used in CEM03.01
and LAQGSM03.01 with the fission-like binary-decay model GEMINI
of Charity {\it et al.} \cite{GEMINI}--\cite{Charity01}
which considers production of all possible fragments. 
This way, we have produced the ``G" version of our codes 
(``G" stands for GEMINI),
CEM03.G1 and LAQGSM03.G1. 

The INC, preequilibrium, Fermi break-up, evaporation, fission,
and coalescence
models used in the current versions of our codes are
described in detail in \cite{PhotoCEM03,LAQGSM,ResNote05,CEM03.01}
and references therein,  while SMM and GEMINI 
incorporated into the ``S" and ``G" versions,
are described in the original publications
\cite{SMM}--\cite{SMM_history} and 
\cite{GEMINI}--\cite{Charity01}, respectively.

{\bf 2. Results}

We have incorporated into CEM03.01 and LAQGSM03.01 GEMINI 
and SMM as provided to us by their authors, Prof.\ Charity
and Dr.\ Botvina, without any essential changes to or fitting of 
their parameters.
A few ``cosmetic" changes were made only to accommodate
them to our FORTRAN compilers and to fix several observed 
``bugs".

The upper-left plot in Fig.\ 1 shows the recent experimental data
\cite{PEPANLett04}
on the mass-number distribution of the product yield from the
reaction 660 MeV p + $^{129}$I compared with results by the standard
version of our CEM03.01 event generator,
as well as with results by our new ``S" and ``G" codes
(similar results are obtained for this reaction with 
LAQGSM03.01 and its ``S" and ``G" versions).
One can see that the standard versions
do not describe production of isotopes with mass number 
$26 < A < 63$ from this reaction
observed in the experiment \cite{PEPANLett04}. These
products are too heavy to be evaporated from compound nuclei
and the target is too light to fission, producing these isotopes
as fission fragments (CEM03.01 and LAQGSM03.01 consider
only ``conventional" fission of preactinides and actinides
and do not consider at all fission of nuclei with $Z < 65$).

The ``S" and ``G" versions do predict such isotopes and
agree reasonably well with available experimental data. 
This is the main reason we have developed the ``S" and ``G" versions
of our codes.  The results by the ``S" version for the  
$A$-distribution of product yield are very similar to the ones
from the ``G" version for the entire range of product masses,
except the region of light fragments $10 < A < 20$, 
where there are no experimental data. 
From this plot we see only that 
products with  $26 < A < 63$ are produced in this reaction
and they can be described either via fission-like
binary decays (the ``G" versions of our codes), or as
products of multifragmentation of highly-excited nuclei
(the ``S" versions), without a distinctive preference.

\clearpage            

\begin{figure}[ht]                                                 

\centering
\includegraphics[height=175mm,angle=-0]{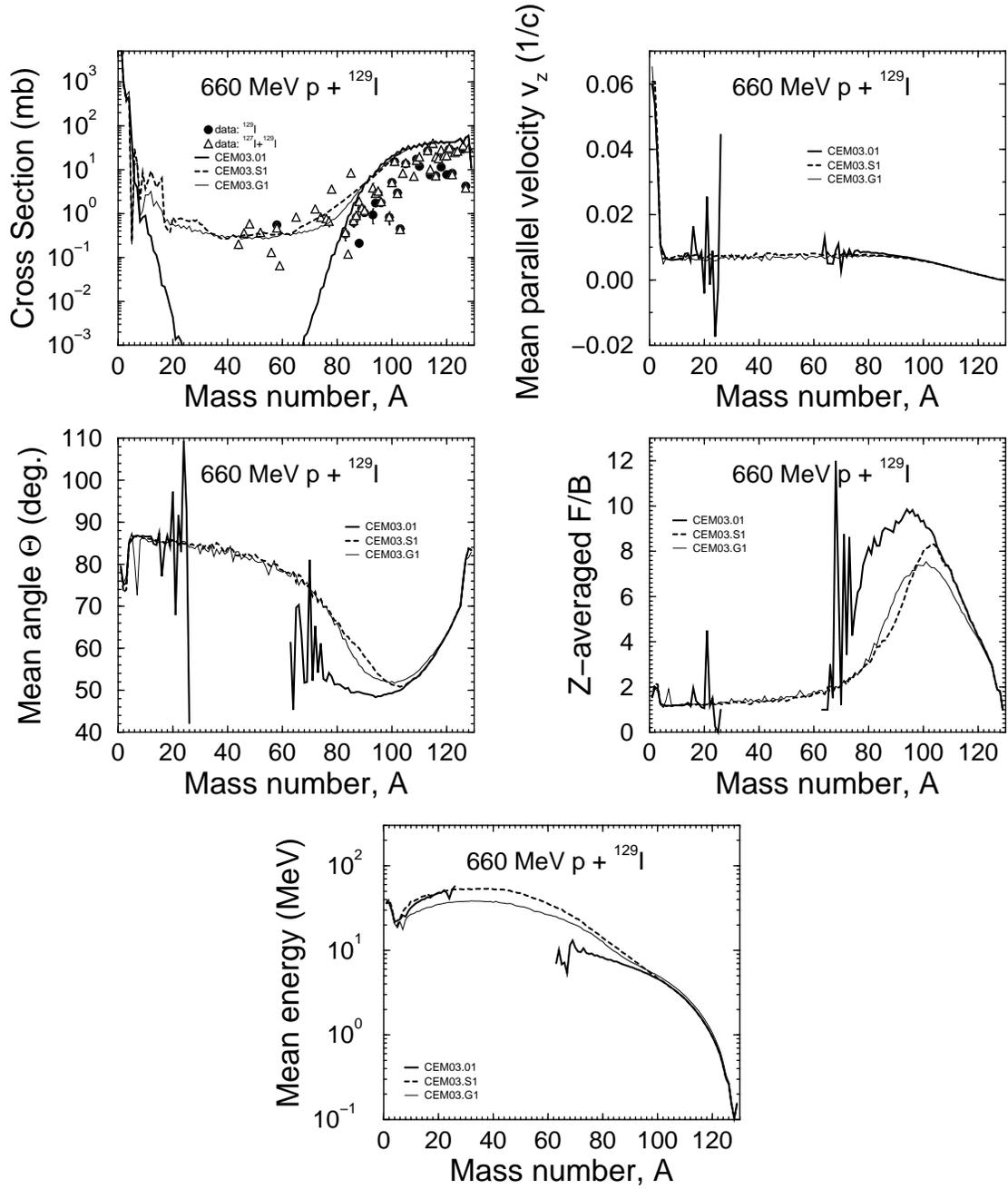}

\caption{Predictions of CEM03.01, CEM03.S1, and CEM03.G1 for the mass
number distribution of the product yield,
mean parallel velocity $v_z$, mean production angle $\Theta$,
Z-averaged A-dependence of the F/B ratio of the 
forward product cross sections to the backward ones,
and the mean kinetic energy of all products in the laboratory system
for the reaction 660 MeV p + $^{129}$I (lines)
compared with available experimental data (symbols) 
\cite{PEPANLett04},
as indicated. The big fluctuations in the
values by CEM03.01 of $<v_z>$, $<\Theta>$, $<R=F/B>$, and $<T_{kin}>$ 
for masses around $A = 20$ and 65 do not provide real physical information,
as they are related to the limited statistics of our 
Monte-Carlo simulation caused by the very low yield of isotopes 
at the border between spallation 
and fragmentation, and at that between fragmentation and evaporation
(with no events at all in the fragmentation region, $26 < A < 63$,
neglected by CEM03.01).
Our calculation provides only a few (or even one) 
isotopes of a given $A$ in these mass regions, and mean values
for such events do not have any significance.}
\end{figure}

\clearpage            

Hoping to understand better the mechanisms for the production 
of isotopes with  $26 < A < 63$
and of other products as well, we also calculate with all versions
of our codes several kinematic properties of products:
the mean parallel velocity $<v_z>$,
the mean production angle  $< \Theta >$, 
Z-averaged A-dependence of the F/B ratio of the 
forward product cross sections to the backward ones $< R >$,
and the mean kinetic energy of all products in the laboratory system,
$<T_{kin}>$,
showed in other subplots of Fig.\ 1.
Such characteristics can be measured with some specific techniques
for some nuclear reactions and have proven to be very useful in understanding 
reaction mechanisms \cite{PhotoCEM03}, although the activation
technique of the experiment \cite{PEPANLett04} does not provide
such measurements. There are big differences among results 
provided by the standard version CEM03.01, and the ``S" and ``G" versions
for  $<\Theta>$, $<R>$, and  $<T_{kin}>$
for products with $70 < A < 100$. There is also
a significant difference between predictions by the ``S" and ``G" version for
 $<\Theta>$ of products with $85 < A < 110$,
for $<R>$ of the same products,
and for $<T_{kin}>$ of products with $15 < A < 80$.
Unfortunately, none of these characteristics have been measured, so
we cannot choose a specific reaction mechanism
based on these results.

Fig.\ 2 shows results similar to the ones presented in Fig.\ 1,
only for a lighter target, $^{56}$Fe, and a lower
energy of 300 MeV measured in inverse kinematics at GSI 
as 300 MeV/nucleon $^{56}$Fe + p \cite{Villagrasa,Carmen}.
The situation with the agreement or disagreement of our calculations with
the data  \cite{Villagrasa,Carmen} is very similar to that shown
in Fig.\ 1, and all comments here are the same. The standard versions
of our event generators strongly underestimate production of fragments with 
$A <32$ from this reaction. These fragments can be described either
via fission-like binary decays (the ``G" versions of our codes), or as
products of multifragmentation of highly-excited nuclei
(the ``S" versions). Comparing only the total production
cross sections with experimental data (Fig.\ 2, for $Z$-integrated
$A$-dependence of the yield,  and Fig.\ 3, where we show
cross sections of all measured isotopes, separately)
does not allow us to identify the ``real" nuclear reaction mechanisms for
the production of these isotopes. 
Kinematic properties of products discussed above, like 
$< \Theta >$, $< R >$, and $<T_{kin}>$ are different
for different models and could shed more light on the 
mechanisms of these nuclear reactions, but such characteristics were
also not measured in this experiment
 \cite{Villagrasa,Carmen}. In addition to the production
cross section, the GSI inverse-kinematics technique
provides also the mean parallel velocity $v_z$ of all products in the
reference frame of the projectile. Referring to the upper-right
subplot of Fig.\ 2, $<v_z>$ is not sensitive enough to the reaction
mechanisms, and all three versions of our codes, ``S", ``G", and
the standard version ``03.01" provide almost the same $<v_z>$,
for this particular reaction. 
The mean kinetic energy of products is more sensitive to
the reaction mechanisms considered, therefore more informative.
As one can expect in advance, the multifragmentation mechanism
(``S" version of our codes)
provides more energetic light fragments (see the upper plots in Fig.\ 4,
for $Z = 3$, 6, 9, 12, and 15) than the fission-like
binary decay model GEMINI (``G" version)
and the ``conventional" evaporation model considered 
by our standard ``03.01" version do. With increasing
mass/charge of the products, this difference diminishes, and for
isotopes with $Z \ge 20$, all three versions of our codes predict
the same values of $<T_{kin}>$ (lower plots in Fig.\ 4).
Unfortunately, we do not have experimental data for
these quantities, and therefore are not able to
make an unambiguous choice between multifragmentation and
binary decays in the production of light fragments from this reaction.
\clearpage      

\begin{figure}[ht]                                                 

\centering
\includegraphics[height=200mm,angle=-0]{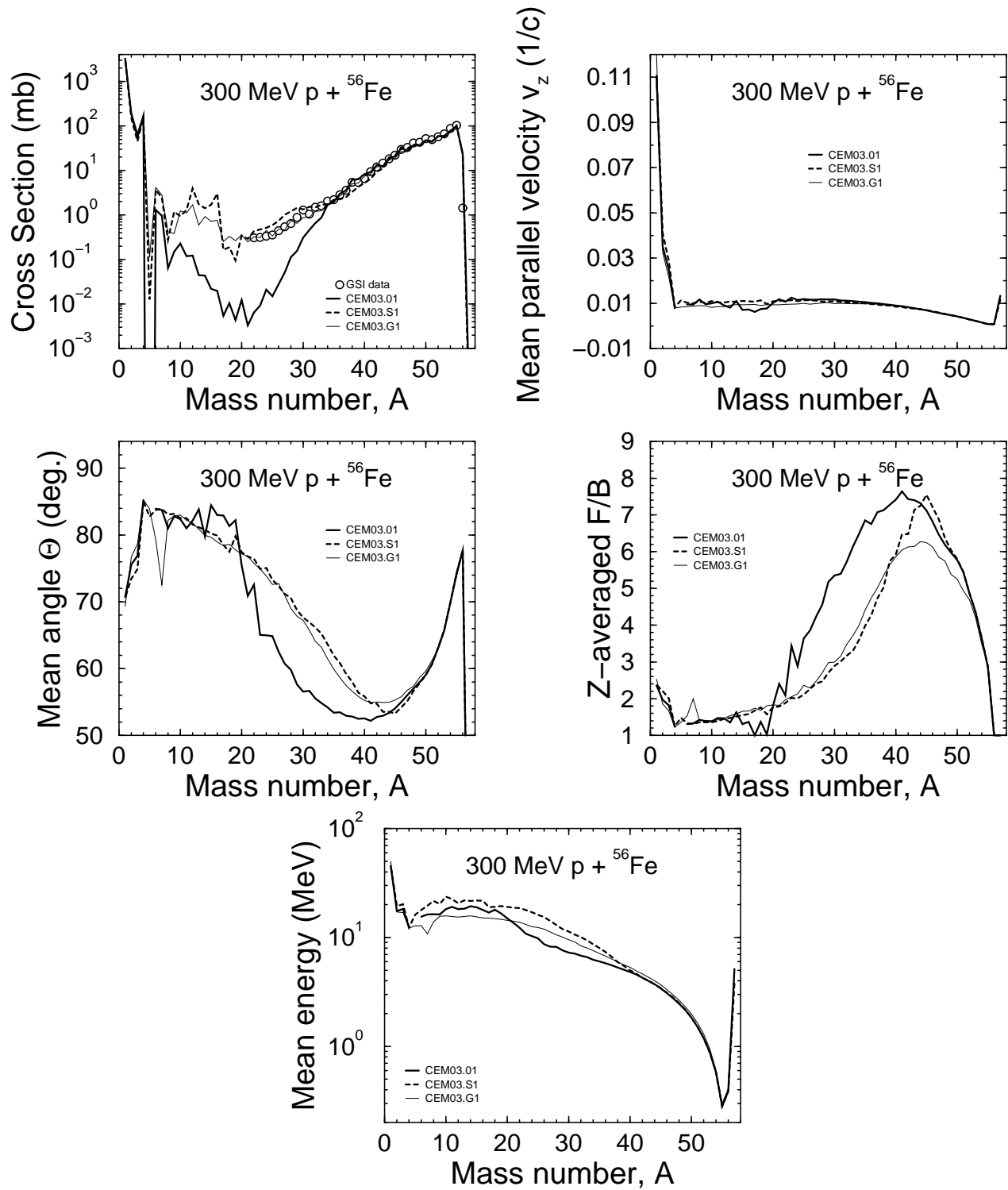}
\caption{The same as Fig.\ 1, but for the reaction
300 MeV p + $^{56}$Fe measured at GSI in inverse kinematics 
\cite{Villagrasa,Carmen}.
}
\end{figure}

\begin{figure}[ht]                                                 
\centering
\includegraphics[height=220mm,angle=-0]{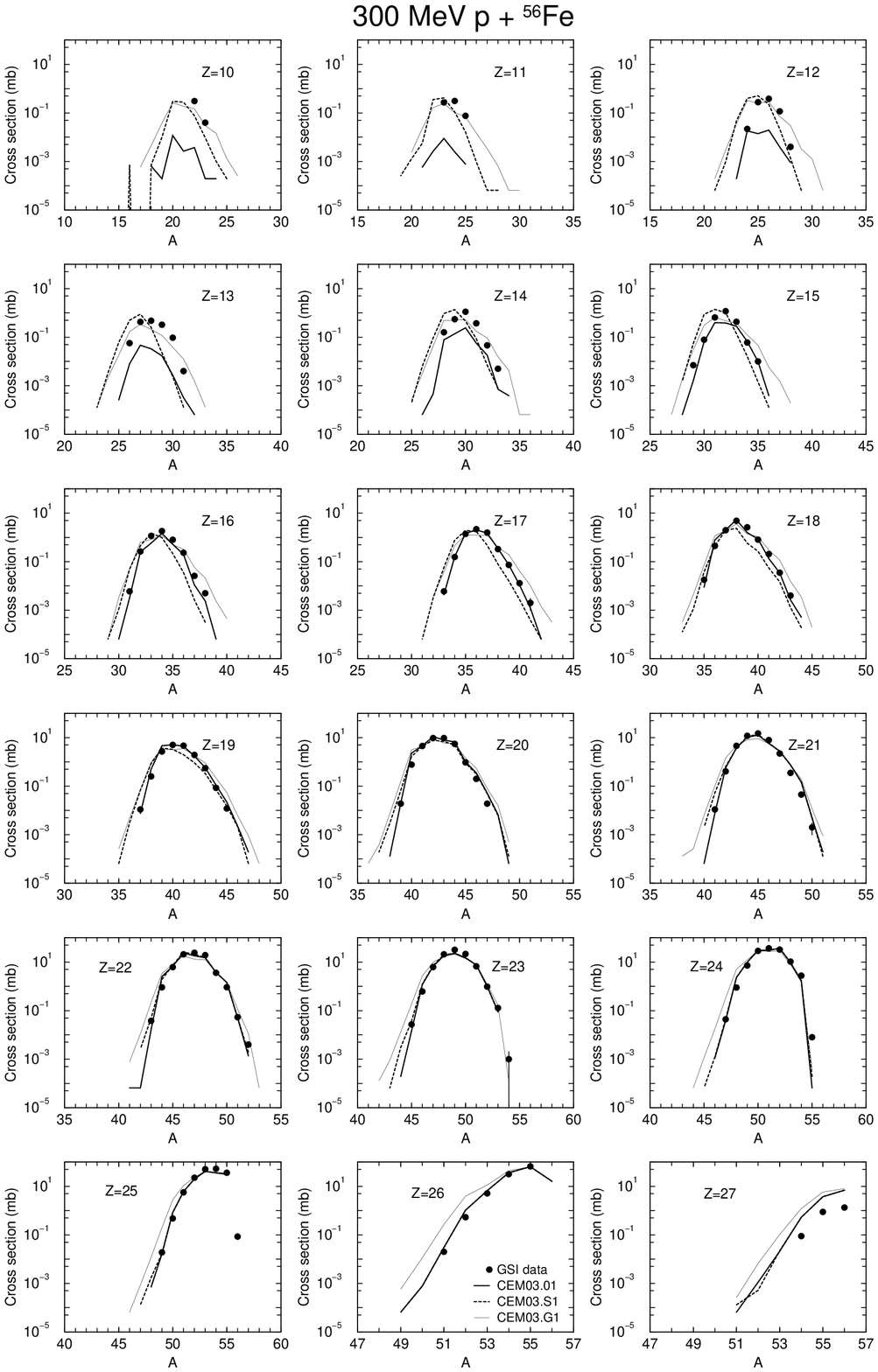}
\caption{Comparison of all measured 
\cite{Villagrasa,Carmen}
cross sections (symbols) of products from the reaction 300 MeV p + $^{56}$Fe
with CEM03.01, CEM03.S1, and CEM03.G1 results (lines), as indicated.}
\end{figure}

\begin{figure}[ht]                                                 
\centering
\includegraphics[height=200mm,angle=-0]{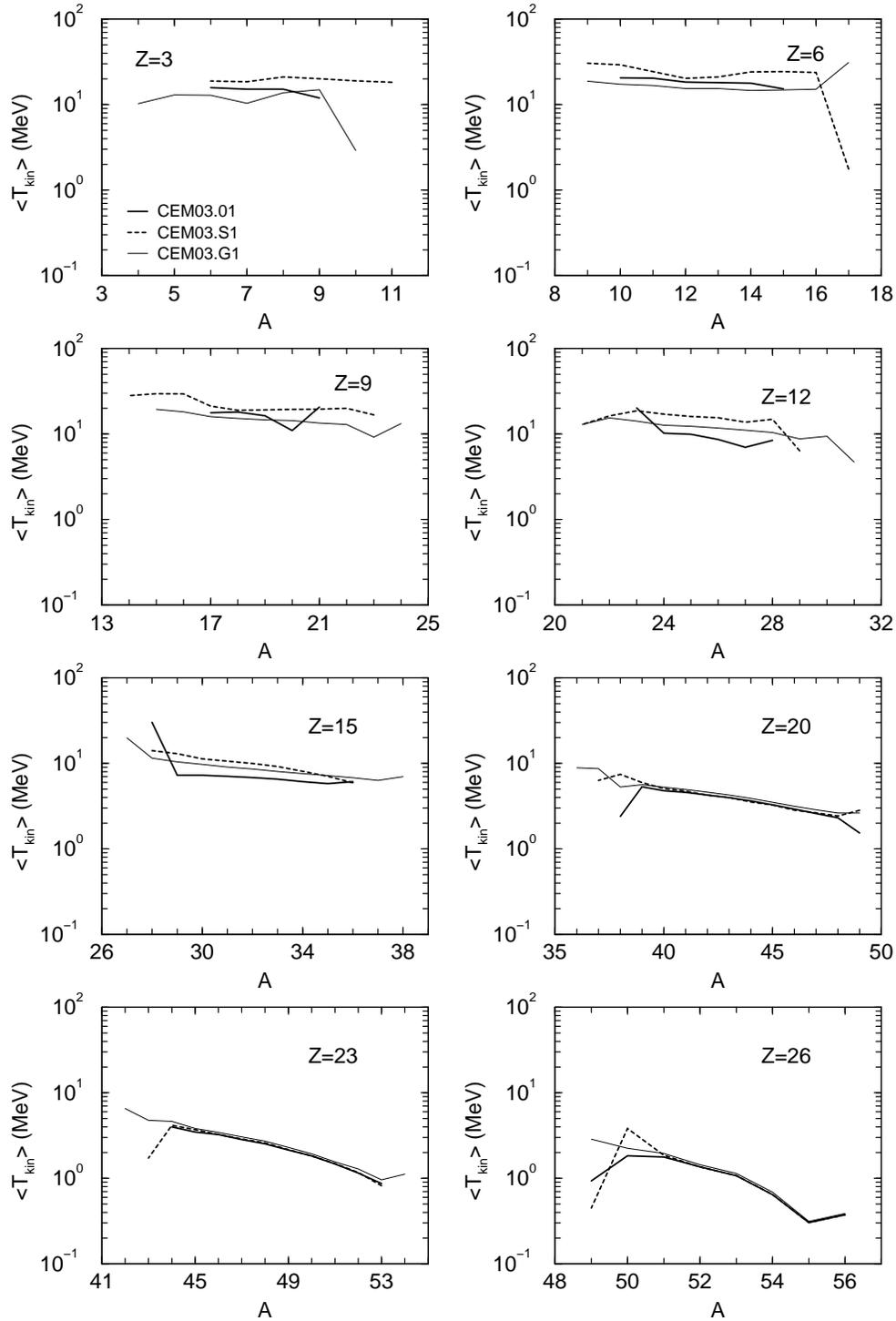}  
\caption{Predictions by CEM03.01, CEM03.S1, and CEM03.G1 for the
mean kinetic energy of eight nuclides with $Z = 3$, 6, 9, 12,
15, 20, 23, and 26 produced in the reaction 300 MeV p + $^{56}$Fe
(no experimental data are available to us).
The big fluctuations in the
values of $<T_{kin}>$ at both ends of distributions
do not provide real physical information,
as they are related to the limited statistics of our 
Monte-Carlo simulation caused by the very low yield of 
extremely neutron-rich and neutron-deficient isotopes. 
Our calculation provides only a few (or even one) 
isotopes of a given $A$ in these regions, and mean values
for such events do not have any significance.}
\end{figure}

\clearpage

Figs.\ 5--7 provide results for the same reaction and are similar
to the ones shown in Figs.\ 2--4,
but for a higher energy of 1 GeV, also measured
in inverse kinematics at GSI by Carmen Villagrasa {\it et al.}
\cite{Villagrasa,Carmen} and by Paolo Napolitani {\it et al.}
\cite{Napolitania,Napolitanib}.
At 1 GeV (Figs.\ 5--7), the situation changes a little
in comparison with what we have above at 300 MeV: 
The energy here is higher and the reaction is deeper,
the target, $^{56}$Fe, is not too heavy, 
so the standard version of our codes
describe reasonably well all the measured product yields.
At 1 GeV, the standard versions of our models predict
light fragment production via deep spallation processes of the INC
followed by preequilibrium and evaporation,  
without considering the multifragmentation (``S" version)
or/and binary-decay processes (``G" version).
It is interesting that at 1 GeV, the standard ``03.01" event generators
describe these cross sections measured at GSI even better
than the ``S" or ``G" versions do, especially for products with 
$5 < A < 16$ (see the upper-left plot in Fig.\ 5) and
$Z < 10$ (see Fig.\ 6). However, we consider this fact
only as natural result of several years of careful 
development of our standard event generators
rather than an indication that no multifragmentation and/or
binary decays occur at 1 GeV (from a physical point of view,
if we have multifragmentation and/or binary decays
at 300 MeV, one may expect to have them even more pronounced
at 1 GeV): Our standard event generators
CEM03.01 and LAQGSM03.01 consider the INC, preequilibrium, evaporation,
and the coalescence models for the production of
isotopes from this reaction, and each of these models have their own 
parameters. These parameters have been adjusted, then fixed,
while developing CEM03.01 and LAQGSM03.01
so that our codes describe as well as possible arbitrary nuclear reactions.
The ``G" and ``S" versions are produced \cite{ResNote06} without any fitting
or adjustment of any parameters. We think that by adjusting
and fitting the parameters of the ``G" and ``S" versions, one might
obtain an agreement of their results for the production
cross sections no worse than that provided by the
standard versions of our codes. A difference would likely be observed 
for predictions of other characteristics of these reactions,
like the kinematic properties of products discussed previously.
In the framework of the versions we have so far,
the biggest difference among results for this reaction
by the ``S", ``G",
and the standard ``03.01" versions is for $<\Theta>$ and $<R>$
for products with $12 < A < 40$, for $<T_{kin}>$ of products
with $5 < A < 30$, with
also quite a big difference in the $Z$-integrated $A$-dependence
of the yield for light fragments with $5 < A < 18$
(see the upper-left subplot in Fig.\ 5).

The GEM2 evaporation/fission model \cite{GEM2}--\cite{Furihata3} 
does not consider the angular momenta of the emitted particles,
therefore the angular momenta of nuclei calculated at the INC 
and preequilibrium stages of reactions
are not used at all and neglected in evaporation and fission
processes. The same is true for the ``S" version of our codes.
On the other hand, GEMINI 
\cite{GEMINI}--\cite{Charity01} does consider angular momenta
of all products, so the ``G" version of our codes can be 
used to study the effect of angular momentum
in nuclear reactions. To reveal the effect of angular momentum, $L$,
of the compound nucleus on the ``recoil characteristics"
$<\Theta>$, $<R>$, and $<v_z>$
of the reaction studied here, we have performed additional calculations
with the ``G" version of our codes assuming angular momentum of all
compound nuclei being equal to zero. Results of such a modification
of CEM03.G1 are shown in Fig.\ 5 with thin dashed lines, to be compared with
the solid thin lines showing results by CEM03.G1 considering the real
angular momenta of all compound nuclei. We see that the effect of angular 
momentum, $L$, of the compound nucleus on results for
$<\Theta>$, $<R>$, and $<v_z>$
calculated by GEMINI in CEM03.G1 is more important for products
with $12 < A < 46$, but is not very strong, on the whole.
\clearpage

\begin{figure}[ht]                                                 
\centering
\hspace*{-10mm}
\includegraphics[height=210mm,angle=-0]{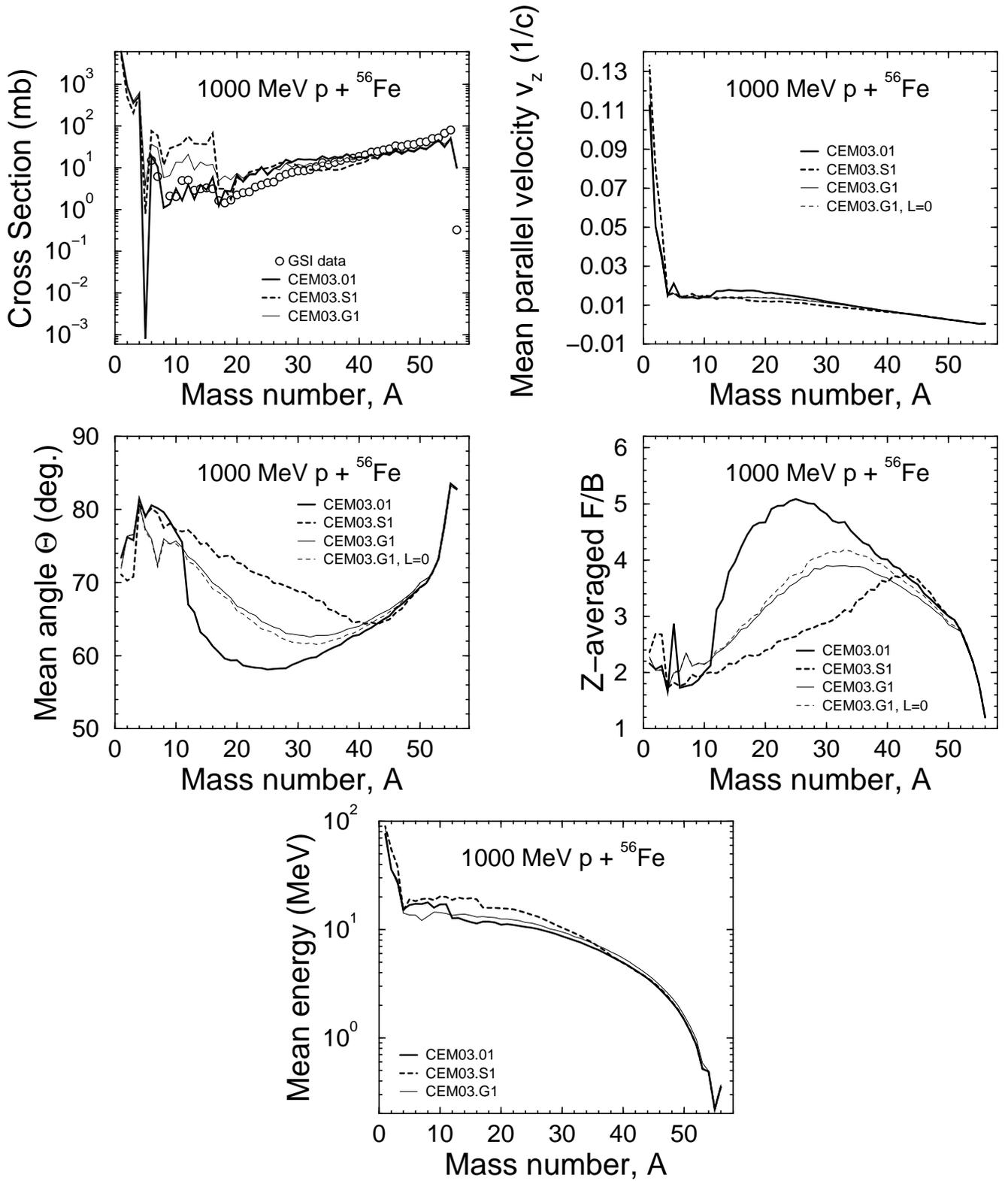}
\caption{The same as Fig.\ 2, but for the reaction
1000 MeV p + $^{56}$Fe measured at GSI in inverse kinematics 
by Villagrasa (medium and heavy products) 
\cite{Villagrasa,Carmen} 
and by Napolitani (light fragments) 
\cite{Napolitania,Napolitanib} 
with coauthors. To reveal the effect of angular momentum, $L$, of the
compound nucleus on results calculated by GEMINI in CEM03.G1,
the dashed thin lines show results obtained assuming $L = 0$ in GEMINI, 
that should be compared with the results shown by thin solid lines obtained
with real values of $L$.}
\end{figure}

\begin{figure}[ht]                                                 
\centering
\hspace*{-5mm}
\includegraphics[height=210mm,angle=-0]{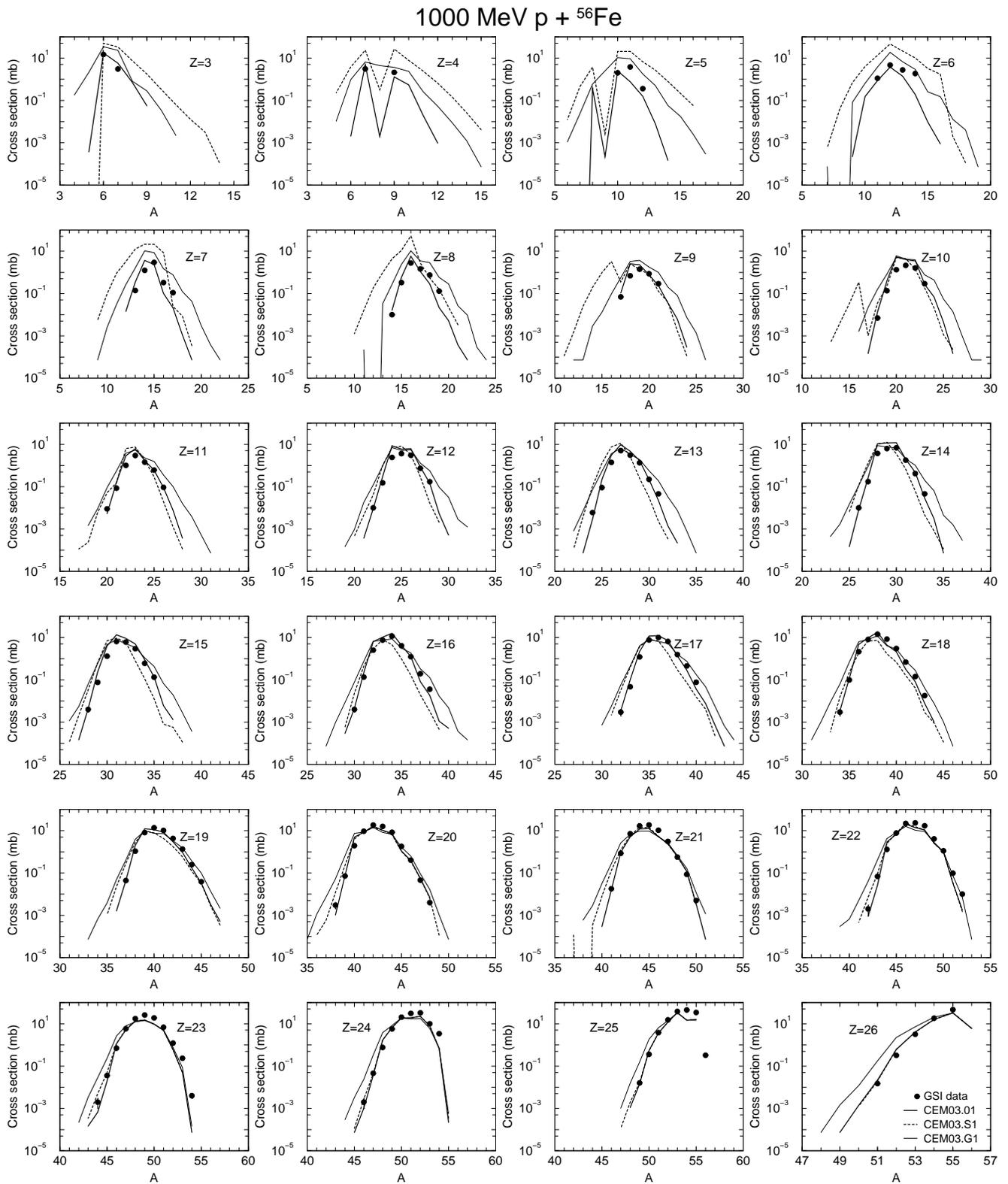}
\caption{The same as in Fig.\ 3, but for the reaction
1000 MeV p + $^{56}$Fe measured at GSI in inverse kinematics 
by Villagrasa (medium and heavy products) 
\cite{Villagrasa,Carmen} 
and by Napolitani (light fragments) 
\cite{Napolitania,Napolitanib} 
with coauthors.}
\end{figure}

\begin{figure}[ht]                                                 
\centering
\includegraphics[height=220mm,angle=-0]{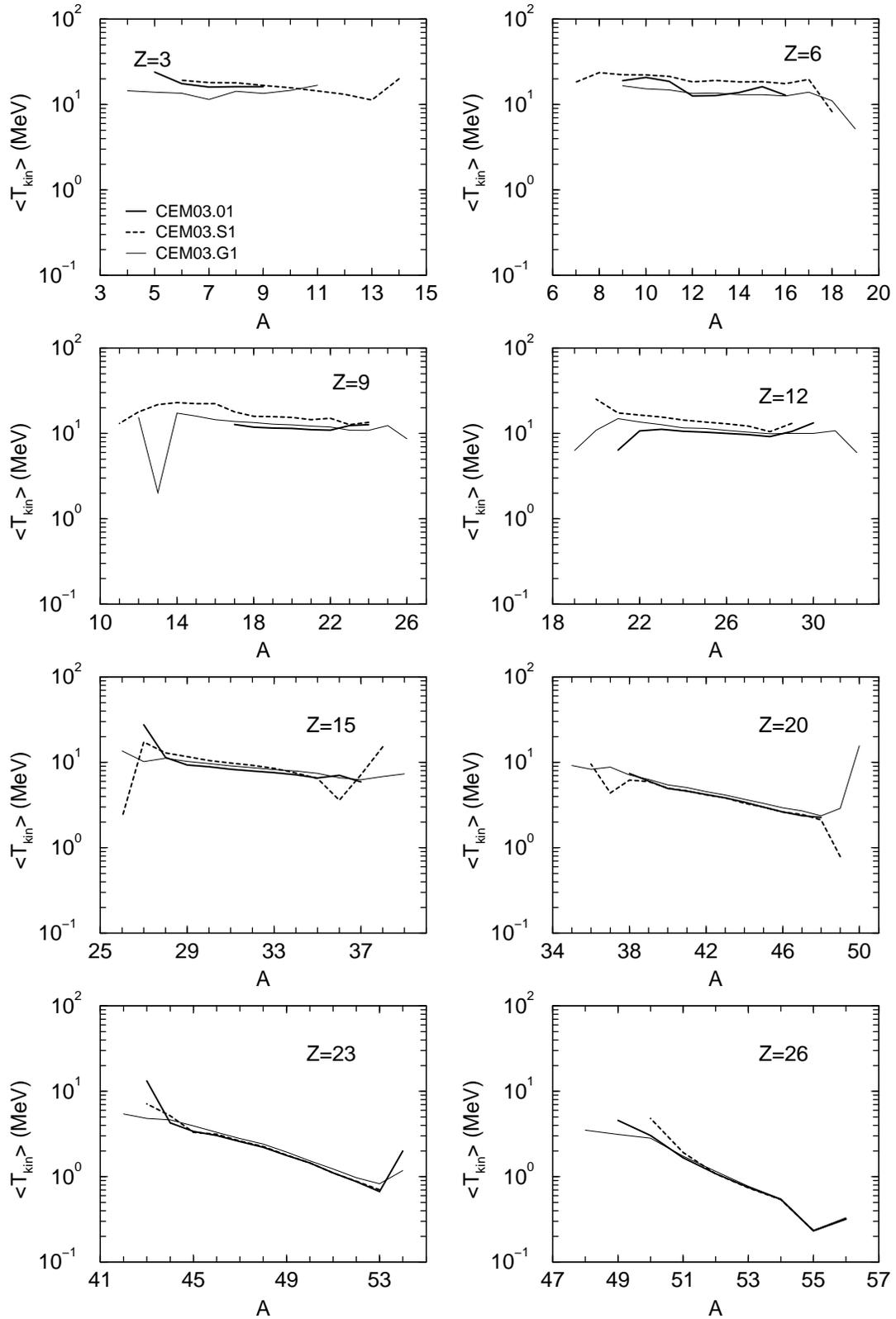}
\caption{The same as Fig.\ 4, but for the reaction
1000 MeV p + $^{56}$Fe.}
\end{figure}

\clearpage

Finally, Fig.\ 8 shows an example of a proton-induced reaction at a
higher energy and a heavier target, namely, 3.65 GeV p + $^{112}$Sn,
measured recently at JINR, Dubna with the activation technique
\cite{Balabekyan06}.
The CEM03.01,  CEM03.S1, and CEM03.G1
results shown in the figure present $A$-distributions of the yield
of all products, {\it i.e.},
sums over $Z$ of yields of all isotopes with a given mass number $A$, 
while the experimental data 
obtained by the activation method generally represent 
results for only some isotopes (sometimes, for only a single isotope) 
that contribute to the corresponding data point.
That is, this comparison is only qualitative, not quantitative and 
provides only an approximate picture of the
agreement between the calculations and measured data
(just as in Fig.\ 1, where the data were also
measured by the activation techniques at JINR).
Activation measurements present the total yield for a given $A$
only for cases when cumulative cross sections that include contributions
from all precursors of all possible $Z$ to the given measured yield;
therefore, in general theoretical calculations 
of $A$-distribution of yields should be higher than
many experimental activation data points. 
A much better, quantitative analysis would be
to compare only the measured cross sections,
isotope-by-isotope. Such a comparison of the measured data with
results by CEM03.01 and LAQGSM03.01 (and by FLUKA and LAHET)
is made in the original publication \cite{Balabekyan06} and is
not an aim of the present work.
As the energy of the reaction shown in Fig.\ 8 is much higher than of all
the other reactions shown in previous figures, the situation is also quite
different. We see that the standard CEM03.01 predicts production of
isotopes with all possible mass numbers, from 1 to 112. Intermediate
isotopes with mass numbers $28 < A < 80$ are produced by CEM03.01
only via deep spallation, {\it i.e.}, the INC, followed by preequilibrium
emission of particles up to $^4$He, followed by evaporation of particles and
light fragments up to $A < 28$ from excited compound nuclei,
without considering multifragmentation and/or binary decays.
The ``S" version considers production of such isotopes also via
multifragmentation, while the ``G" version, via binary decays.
Nevertheless, the yield of products with  $28 < A < 80$ predicted by
the  standard CEM03.01 model is higher than the ones predicted by both
the ``S" and ``G" versions. Only for products with $5 < A < 28$ do
the ``S" and ``G" versions predict a much higher yield than CEM03.01 does.
For fragments with $8 \leq A \leq 16$, the ``G" version predicts a
yield about a factor of five higher than the standard CEM03.01, while
the ``S" version predicts even a higher yield, almost two orders of
magnitude more than  CEM03.01 does. Unfortunately, no experimental
data for such products from this reaction are available presently,
so the question about the  ``real" mechanisms for the production 
of such isotopes and their yields remains open. 

Fig.\ 9 shows one example of proton spectra from 500 MeV p + $^{58}$Ni
calculated with our CEM03.01,  CEM03.S1, and CEM03.G1 codes
compared with experimental data by Roy {\it et al.} \cite{roy81}.
As one may expect in advance, all three versions of our codes
provide very similar results, in a good agreement with
the measurement. The spectra by ``S" and ``G" versions
are a little lower in the energy range $T_p \simeq$ 25--50 MeV
in comparison with the standard version CEM03.01,
but the difference is less than a factor of two, and there are
no experimental data for this part of spectra, so again it is
difficult to conclude which version works better here.

Figs.\ 10 and 11,
compare results by CEM03.01, CEM03.S1, and CEM03.G1 for the
total production cross sections of H, He, Li, and Be isotopes
produced in interactions of 1.2 GeV protons with 13 target nuclei
from Al to Th, measured just recently at the Cooler Synchrotron
Facility COSY of the Forschungszentrum J\"{u}lich \cite{Herbach06}.
The experimental data shown in Figs.\ 10 and 11 are taken
from Tables 4, 5, and 6 of Ref. \cite{Herbach06} and present the
measured production cross sections of H, He,  Li, and Be isotopes
only with kinetic energies below 100 MeV. We do not modify 
our codes to
account for this experimental upper limit
of the energy  of detected 

\begin{figure}[ht]                                                 
\centering
\hspace*{-5mm}
\includegraphics[height=220mm,angle=-0]{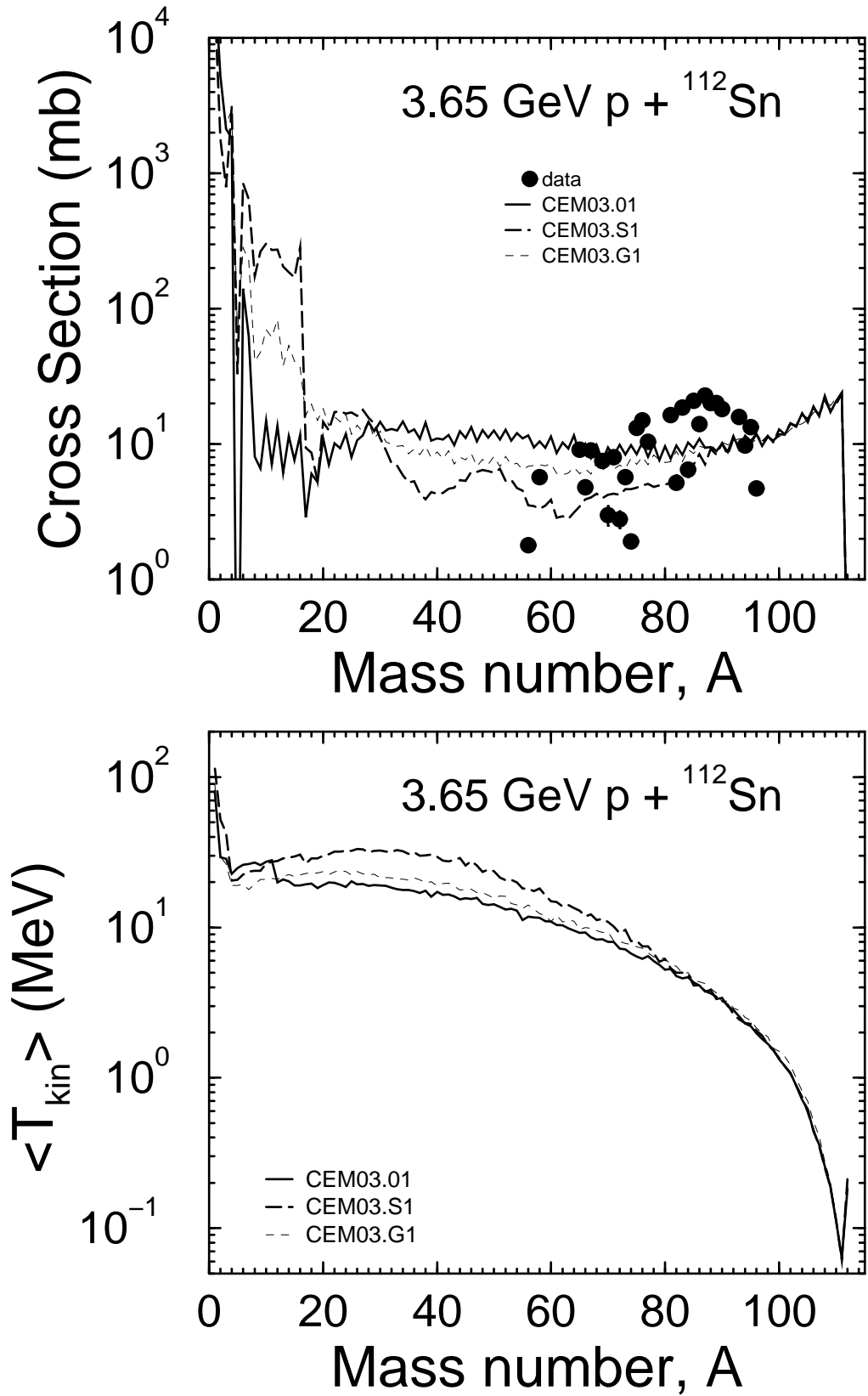}
\caption{Predictions by CEM03.01, CEM03.S1, and CEM03.G1 for the mass
number distribution of the product yield
and the mean kinetic energy of all products in the laboratory system
for the reaction 3.65 GeV p + $^{112}$Sn
 (lines)
compared with available experimental data (circles) 
\cite{Balabekyan06}, 
as indicated.
}
\end{figure}

\clearpage

\vspace*{5mm}

\begin{figure}[ht]                                                 
\centering
\hspace*{-5mm}
\includegraphics[height=175mm,angle=-90]{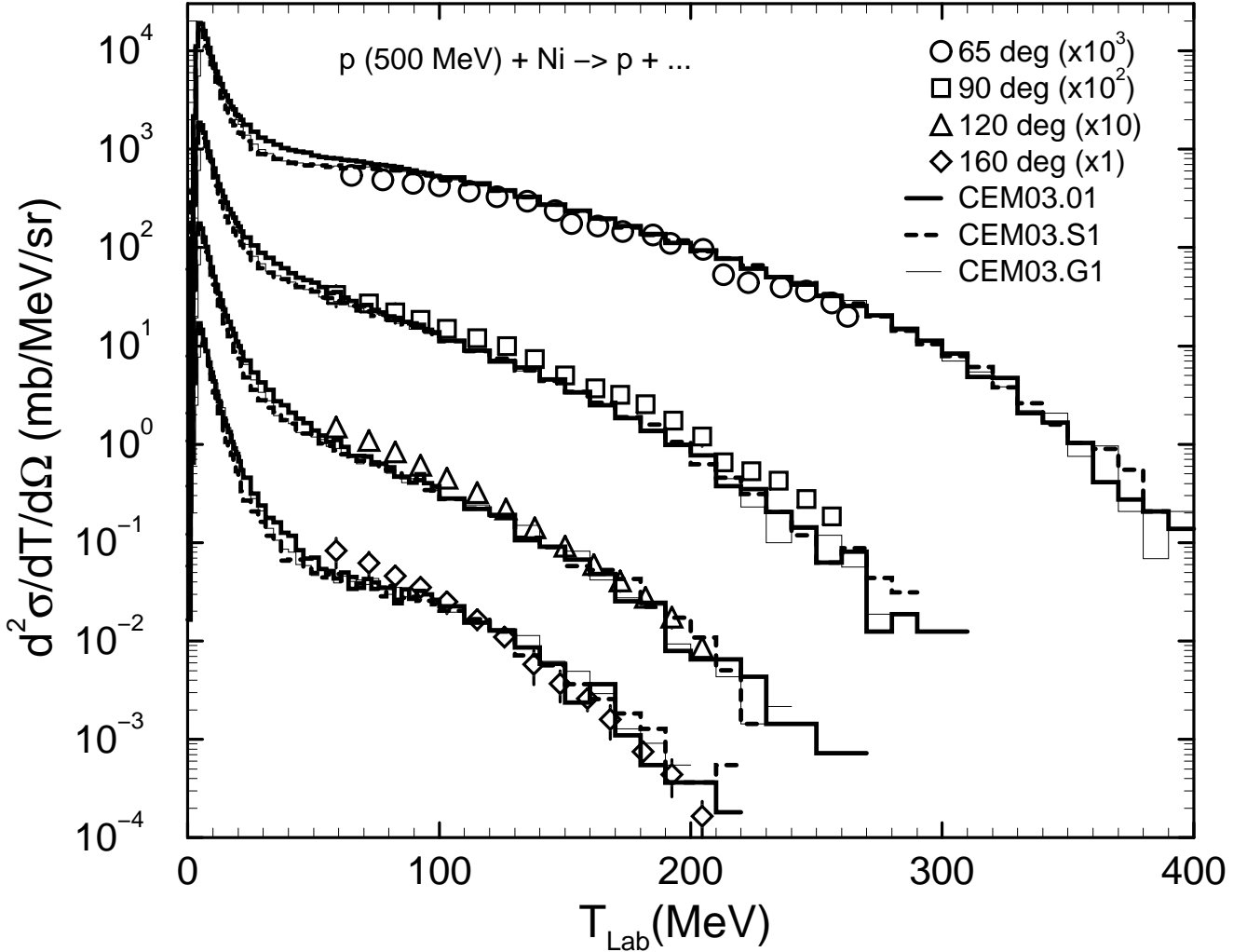}
\caption{
Experimental proton spectra from 500 MeV p + Ni 
\cite{roy81}
compared with CEM03.01, CEM03.S1, and CEM03.G1 results, as indicated.}
\end{figure}

{\noindent
 particles; instead, we estimate 
the contributions from the high-energy tails ($T > 100$ MeV) 
of calculated spectra to the total calculated production cross
sections. 
}
Fig.\ 12 shows an example of our estimates from angle-integrated
energy spectra of p, d, t, $^3$He, and $^4$He calculated
by CEM03.01 for the reaction 1.2 GeV p + Ag. The legend of this figure
presents integrals of spectra (in mb) over the energy for
particle energies above
and below 100 MeV, respectively, and the percentage of contribution
from high-energy tails ($T > 100$ MeV) of spectra to the total
calculated cross sections. We see that for this particular reaction
these contributions are rather small, of only  
about 17\% for p, 3\% for d, 1\% for t, 
2\% for $^3$He, and 0.4\% for $^4$He. Of course, for other targets
and code versions, these contributions differ,
but on the whole they remain small, limited to a few percent. 
This is why
we can compare in Figs.\ 10 and 11 our total production cross sections
calculated for all energies with experimental data that include
energies only below 100 MeV. 

\vspace*{-5mm}
\clearpage

\begin{figure}[ht]                                                 
\vspace*{3mm}
\centering
\hspace*{-5mm}
\includegraphics[height=92mm,angle=-0]{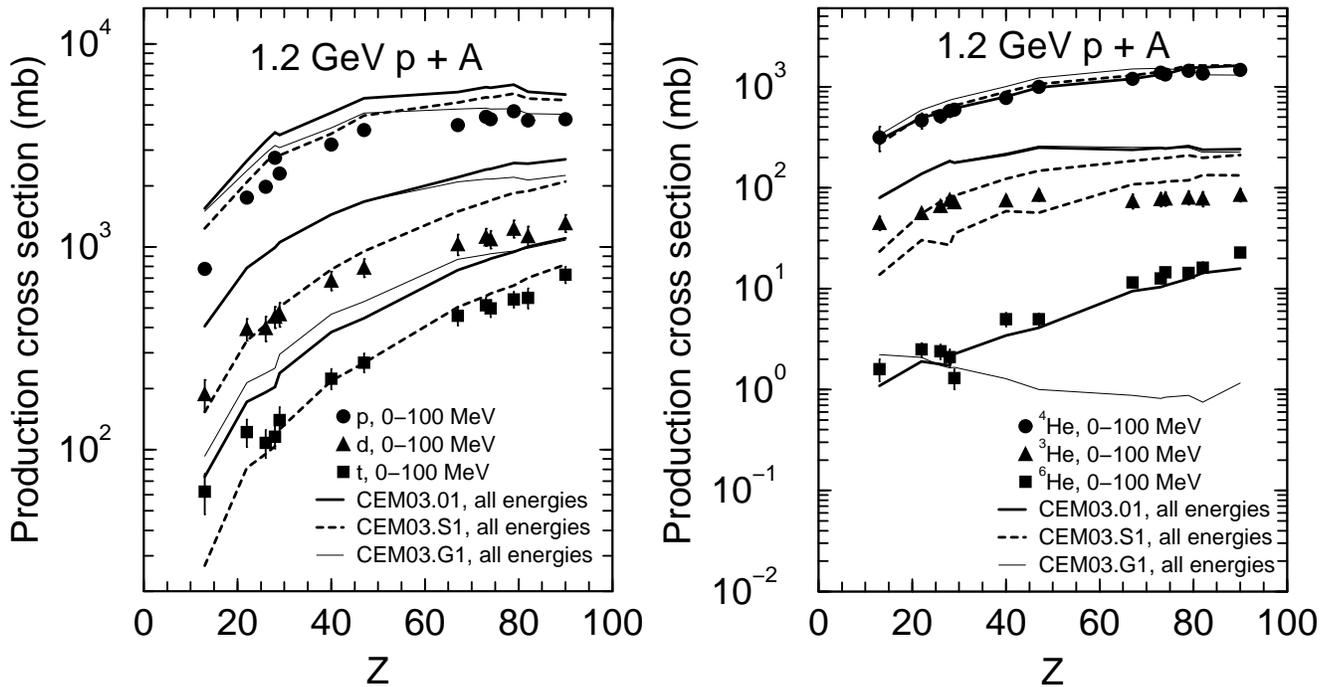}
\caption{Comparison of measured 
\cite{Herbach06}
(symbols)
production cross sections of hydrogen 
and helium isotopes with kinetic energies
below 100 MeV for 1.2 GeV proton-induced reactions
on tirteen targets between Al and Th with results
from CEM03.01, CEM03.S1, and CEM03.G1 (lines), as indicated.
Calculated results include contribution from products of
all possible energies. An estimate of contribution
from high-energy ($T > 100$ MeV) 
tails of calculated spectra to the total calculated production yields
is shown in Fig.\ 12:
It is of ounly about 17\% for p, 3\% for d, 1\% for t, 
2\% for He3, and 0.4\% for He4, in the case of Ag, using CEM03.01.}
\end{figure}

On the whole, with only a few exceptions,
all versions of our codes describe
reasonably the shape and the absolute values of the
measured total production cross sections for all particles,
from protons to $^{10}$Be. The ``S" version overestimates
by several times the yields of all Li isotopes and of
$^{10}$Be from light and medium
nuclei, the yields of $^7$Be and $^9$Be from all targets,
and up to an order of magnitude the yields of $^6$He
from all targets. On the other hand, it agrees better than other
versions of our codes with
the data for all H as well as for  $^3$He and $^4$He isotopes.
The ``G" version predicts reasonably well the yields of all H, all Be,
$^3$He, $^4$He, $^6$Li,  $^7$Li, and not so well for $^9$Li isotopes,
but the shape of the calculated lines for the yields of $^8$Li,
and especially of $^6$He, disagrees with the data.
On the whole, a better agreement with all measured data is
observed for the standard version of our code, CEM03.01.

We now switch to analysis of several heavy-ion induced
reactions with different versions of LAQGSM (CEM does not
describe reactions induced by nuclei). 
Figs.\ 13 and 14 show a comparison 
of LAQGSM03.01, LAQGSM03.S1, and LAQGSM03.G1
results for the total production cross sections (yields) of 
nuclides with $Z$ from 10 to 55
(all measured isotopes)
produced from the fragmentation of $^{124}$Xe
in 1 GeV/A $^{124}$Xe + $^{208}$Pb collisions with the 
very recent GSI measurements \cite{Henzlova05}.
Fig.\ 15 shows predictions
for the mass-number distribution of the product yield and  
the mean kinetic energy (in the projectile frame
of reference) of all products from the same reaction
compared with available data \cite{Henzlova05}.

\begin{figure}[ht]                                                 

\vspace*{5mm}
\centering
\hspace*{-5mm}
\includegraphics[height=92mm,angle=-0]{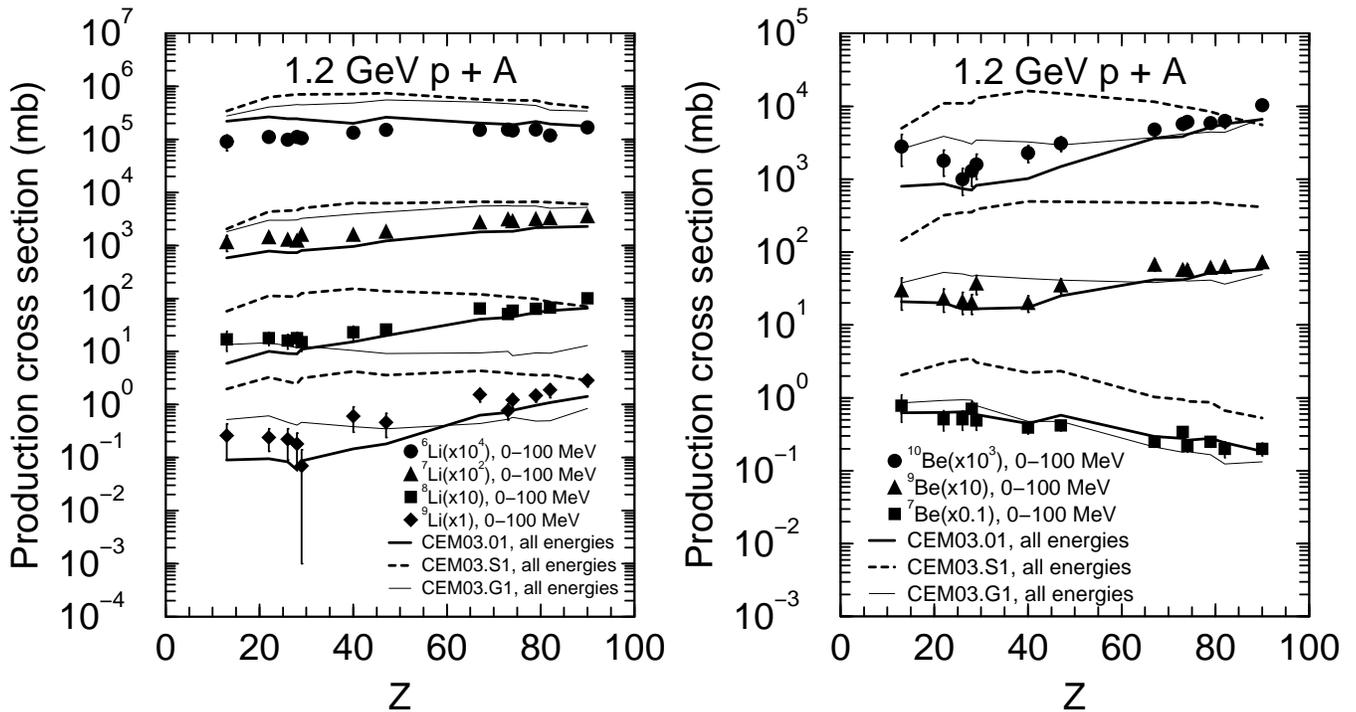}
\caption{The same as Fig.\ 10, but for the production of
Li and Be isotopes.}
\end{figure}

All versions of LAQGSM03.01 describe reasonably well 
cross sections for the production of all measured isotopes,
from Neon to Cesium. A prediction by LAQGSM03.S1
of several unstable Neon isotopes with mass numbers lower than
18 not measured in the experiment (upper-left plot in Fig.\ 13) 
does not bother us much; these unstable isotopes should be disintegrated
into stable nuclei. The transport codes using our event generators 
do take care of this;  we could also add to the codes a check of such
unstable products and disintegrate them
before transferring their results to transport codes.

As observed above for proton-induced reactions, the standard version
LAQGSM03.01 describes the production cross sections of all isotopes
from 1 GeV/$A$ $^{124}$Xe + Pb on the whole a little better than
the ``S" and/or ``G" versions do. We believe that the reason for this
is the same as we had for proton-induced reaction: LAQGSM03.01
was developed carefully for several years; the nuclear
reaction models incorporated into it were adjusted to each other and
their parameters were fitted to describe as
well as possible arbitrary reactions. The ``S" and ``G" versions
of LAQGSM03.01 were developed \cite{ResNote06} 
without any additional fitting or adjustment of any parameters.
It would be possible to adjust the models of
the ``G" and ``S" versions and to fit their parameters so that they
describe production cross sections probably no worse than the standard 
version does, but this is outside the aim of the present work.

From comparison of only the measured \cite{Henzlova05} product
yields with calculations by different versions of LAQGSM it
is difficult, if not impossible, to uncover the ``real" mechanisms
of nuclear reactions contributing to the production of measured isotopes.
In the upper plot of Fig.\ 15, we see a big difference between predictions
by the standard, ``S", and ``G" versions for the yields of isotopes
with $15 < A < 31$, up to an order of magnitude and higher, but,
unfortunately, these products were not measured \cite{Henzlova05}.
We see also quite a big difference between the \\

{\noindent
predictions by
different versions for the mean kinetic energy of products
with $20 < A < 80$  (lower subplot in Fig.\ 15), but
we do not have experimental data for this quantity either.
}

\begin{figure}[h]                                                 
\vspace*{-5mm}
\centering
\hspace*{-15mm}
\includegraphics[height=155mm,angle=-0]{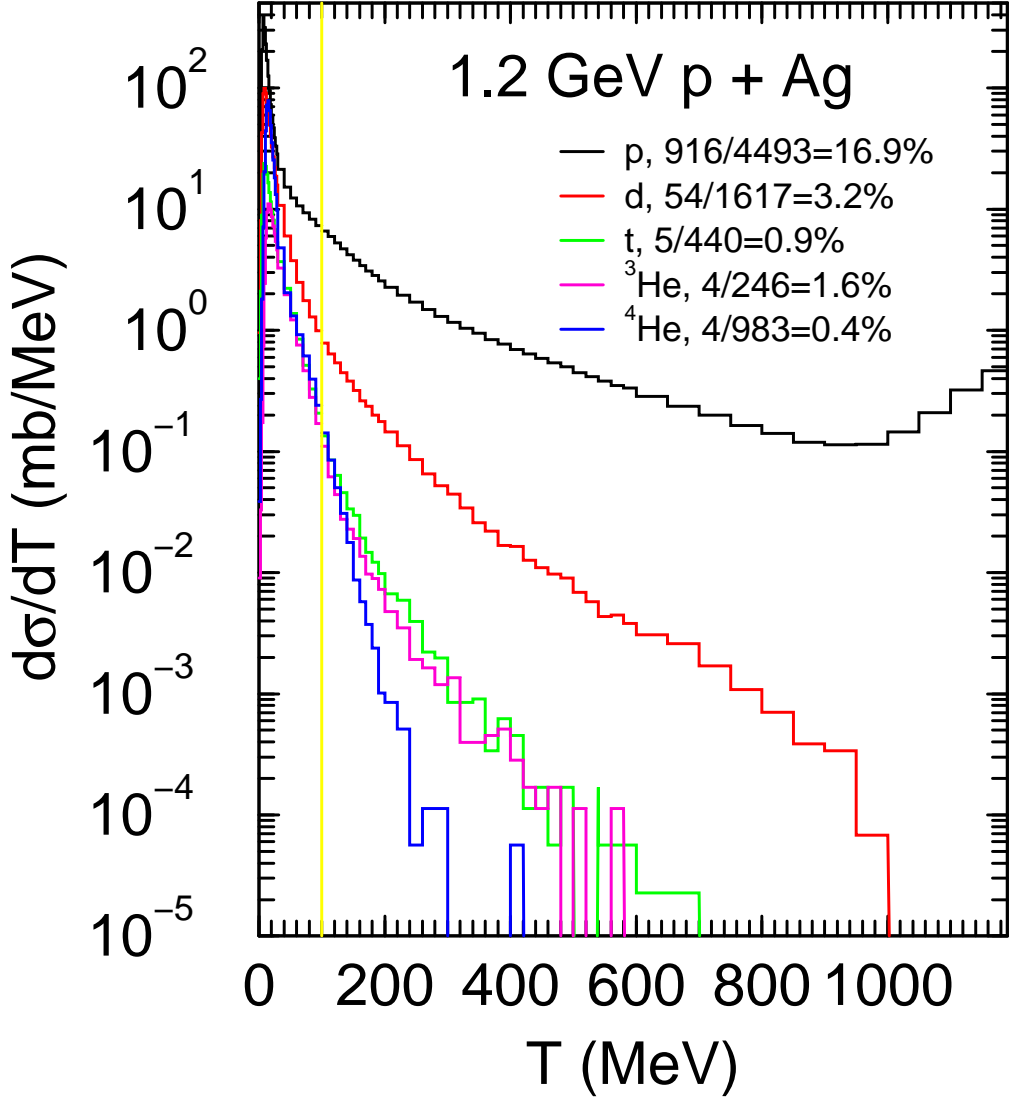}

\vspace*{-5mm}
\caption{Angle-integrated energy spectra of p, d, t, $^3$He, and $^4$He
emitted from the reaction 1.2 GeV p + Ag, as predicted by CEM03.01.
The legend provides integrated production cross sections (in mb) for
particles with energies above and below 100 MeV, respectively.
These integrated cross sections are used to estimate the 
contributions from high energy ($T > 100$ MeV) 
tails of calculated spectra to the total
calculated production yields. These contributions are
about 17\% for p, 3\% for d, 1\% for t, 
2\% for $^3$He, and 0.4\% for $^4$He, 
for results by CEM03.01 for a Ag target.}
\end{figure}
\clearpage            

\begin{figure}[ht]                                                 
\centering
\hspace*{-5mm}
\includegraphics[height=210mm,angle=-0]{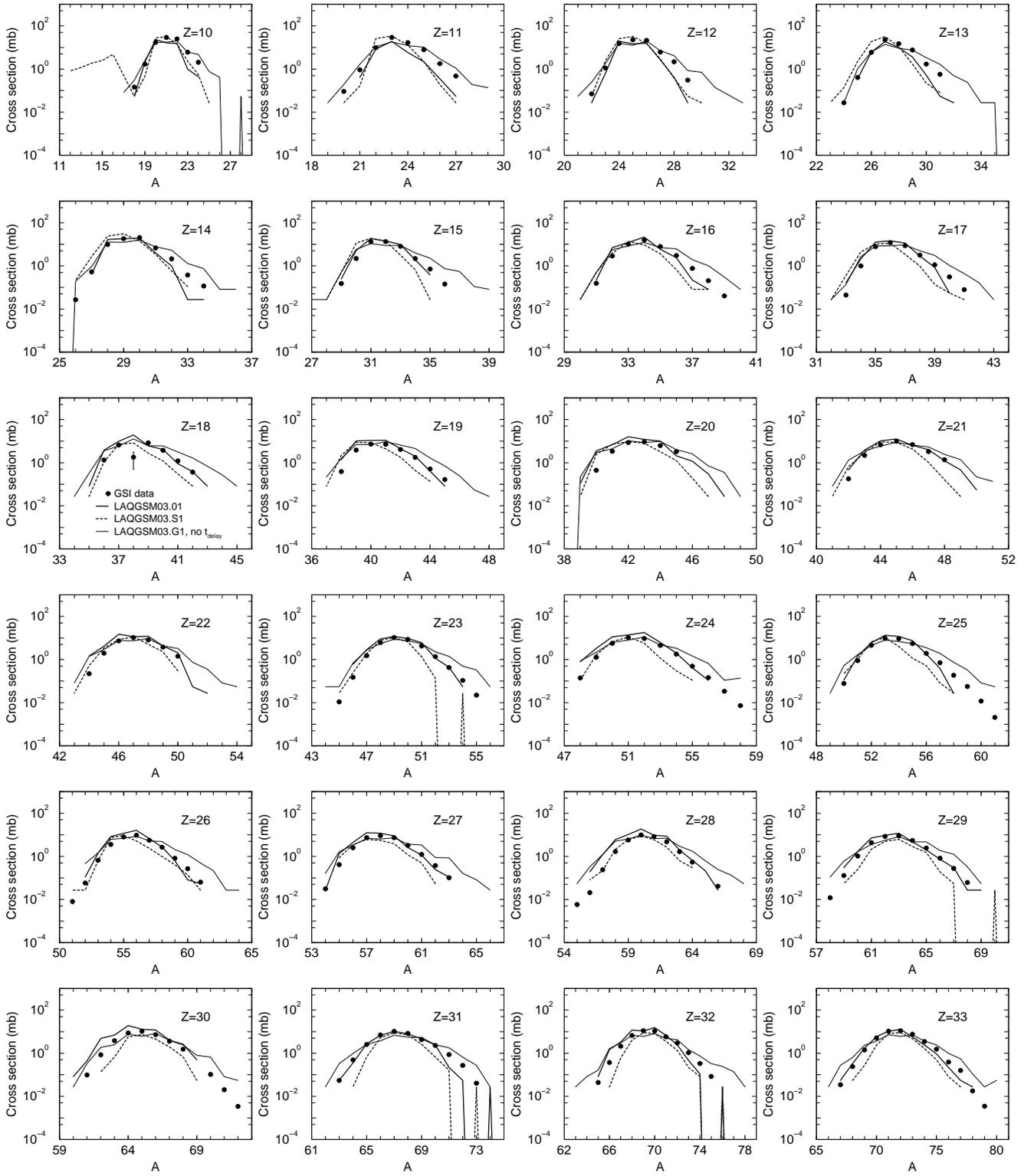}
\caption{Comparison of LAQGSM03.01, LAQGSM03.S1, and LAQGSM03.G1
results (lines) for the total cross sections (yields) of nuclides
with $Z$ from 10 to 33 produced from fragmentation of $^{124}$Xe
in 1 GeV/A $^{124}$Xe + $^{208}$Pb collisions with the recent GSI
measurements 
\cite{Henzlova05}
(circles),
as indicated. No delay time in GEMINI is considered in the LAQGSM03.G1
calculation of this reaction.}
\end{figure}
\clearpage            

\begin{figure}[ht]                                                 
\centering
\hspace*{-5mm}
\includegraphics[height=210mm,angle=-0]{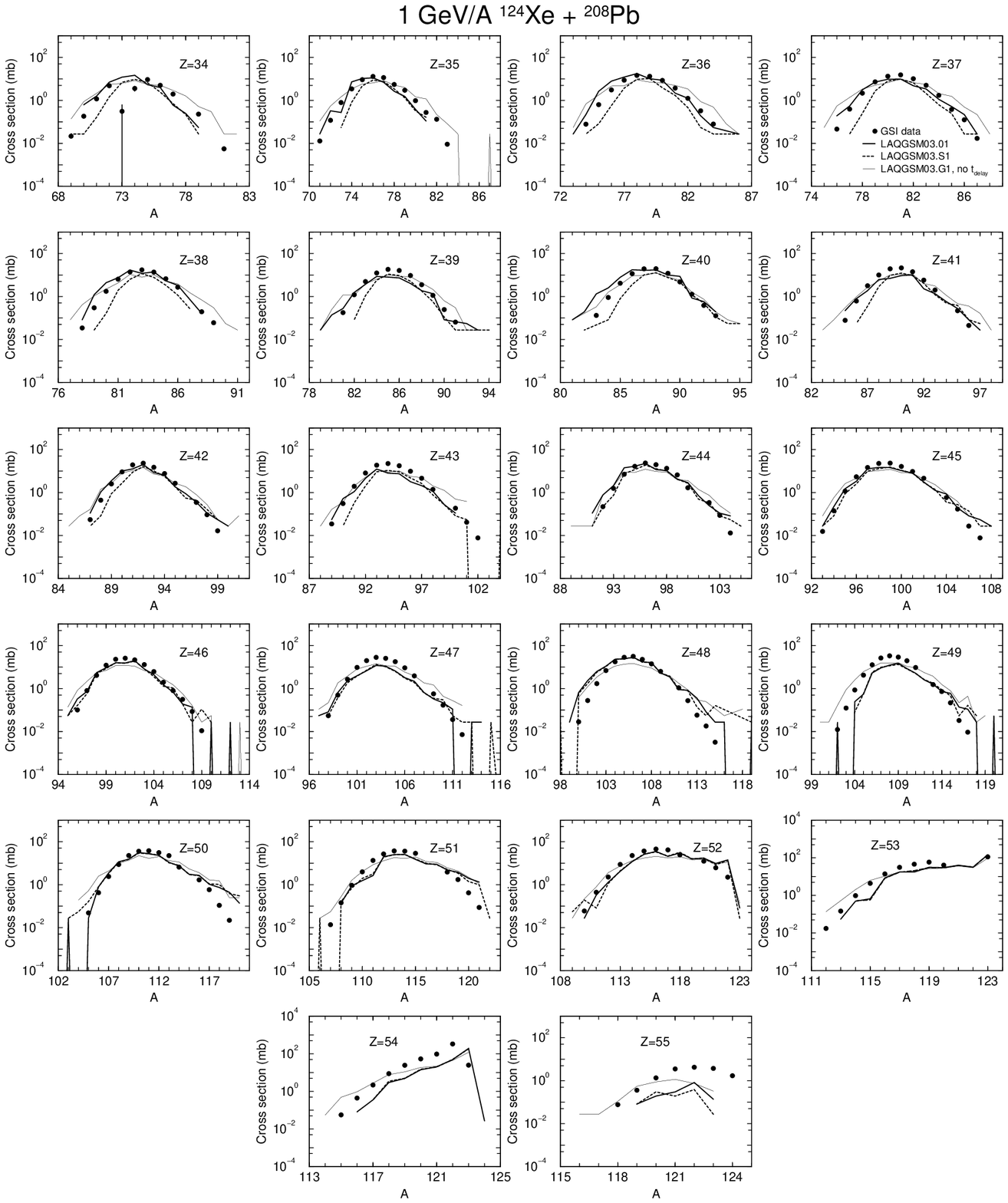}
\caption{The same as Fig.\ 13, but for products with $Z$ from
34 to 55.}
\end{figure}
\clearpage            

\begin{figure}[ht]                                                 
\centering
\hspace*{-15mm}
\includegraphics[height=215mm,angle=-0]{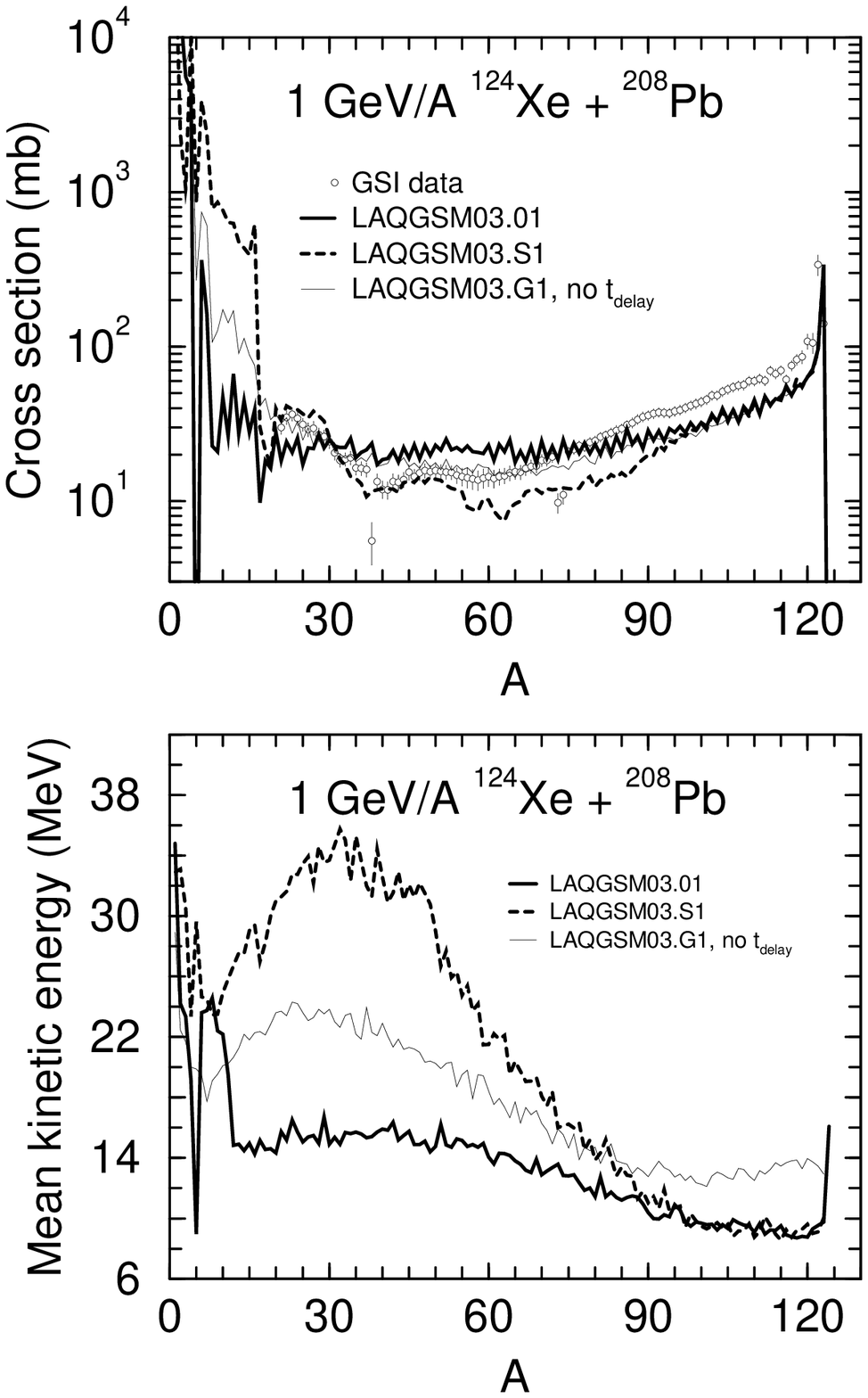}
\caption{Predictions of  LAQGSM03.01, LAQGSM03.S1, and LAQGSM03.G1
for the mass-number distribution of the product yield and  
the mean kinetic energy of all products from the fragmentation of $^{124}$Xe
(in the beam system) from the 
1 GeV/A $^{124}$Xe + $^{208}$Pb reaction (lines)
compared with available experimental data (circles) 
\cite{Henzlova05},
as indicated.}
\end{figure}

\clearpage            

Figs.\ 16 and 17 show results very similar to the ones presented
in Figs.\ 13 and 14, only for another reaction (projectile)
measured lately at
GSI, 1 GeV/$A$ $^{136}$Xe + Pb \cite{Henzlova05} (we make all
calculations on mono-isotopic $^{208}$Pb targets but not on Lead
with a natural composition of isotopes as was measured, 
just as we did for the results presented in Figs.\ 13--15).
The situation for this reaction is very similar to the
one for $^{124}$Xe shown in Figs.\ 13 and 14.
All comments made above for the $^{124}$Xe projectiles (Figs.\ 13 and 14)
are valid and could be repeated here again for reactions
of $^{136}$Xe  (Figs.\ 16 and 17). $^{124}$Xe is the most 
neutron-deficient stable isotope of Xenon, 
while $^{136}$Xe is the most neutron-rich one; this indicates that 
our event generators describe equally well reactions involving
both neutron-deficient and neutron-rich nuclei.

The only difference between results shown in Figs.\ 16 and 17 for $^{136}$Xe
in comparison with results for  $^{124}$Xe shown in Figs.\ 13 and 14
is that for $^{136}$Xe we perform two sets of calculations with the
``G" version of LAQGSM: 1) without taking into account the delay time
(results shown with thin solid lines) and 2) with values
$t_{delay} = 75$ and  $\sigma_{delay} = 50$ 
(results shown with thin dashed lines)
for the time-delay parameters of GEMINI.
The reason for this additional study
for reactions induced by  $^{136}$Xe is to understand how results
by LAQGSM03.G1 depend on the value of the  time-delay parameters of GEMINI:
These parameters 
are considered as input parameters of the model and it is
up to users to chose them. For instance, for proton-induced reactions,
we found \cite{Baznat05}  
that: 1) GEMINI merged with CEM/LAQGSM
provides reasonably good results for medium-heavy targets
without a fission delay time; 2) For preactinides, we have to use
$t_{delay}  = 50$--70 and  $\sigma_{delay} =1$--50, otherwise
GEMINI provides too much fission --- this 
may be related to the calculation of fission barriers of 
preactinides with strong ground-state shell corrections in the 
version of GEMINI we use;
3) The current version of GEMINI does not work well for
actinides.

Our results shown on Figs.\ 16 and 17 (and on the
left panels of Figs.\ 18 and 19)
are only for products of fragmentation of the projectile,  $^{136}$Xe,
just as these reactions are measured at GSI \cite{Henzlova05}.
For such processes,
we do not see a big difference between results by LAQGSM03.G1 obtained
without taking into account the delay time (thin solid lines) and
the ones calculated with $t_{delay} = 75$ and  $\sigma_{delay} = 50$ 
(thin dashed lines). This is similar to what we found for 
proton-induced reactions  \cite{Baznat05}. As  $^{136}$Xe is a  
medium-heavy target, it can be calculated with GEMINI without
taking into account the delay time. The situation changes
dramatically if we look in the laboratory system
at all products from this reaction, just as happens in nature,
produced from both
the projectile  $^{136}$Xe and the target $^{208}$Pb (see the
right panels on Figs.\  18 and 19). 
$^{208}$Pb is a preactinide nucleus and has to be calculated with GEMINI
using  $t_{delay} = 75$ and  $\sigma_{delay} = 50$, according to our
experience gained from studying proton-induced reactions  \cite{Baznat05}
(this is why we choose here these values of  $t_{delay}$ and
 $\sigma_{delay}$). 
From the results presented on plots in the right panels of Figs.\ 18 and 19,
we see that all characteristics of isotopes
produced from the target $^{208}$Pb calculated 
with  $t_{delay} = 75$ and  $\sigma_{delay} = 50$ differ significantly
from the ones calculated without taking into account the delay time
in GEMINI. Unfortunately, these characteristics can not be measured
with the GSI technique, and we have no experimental
data with which to compare our results.

Just as for reactions induced by $^{124}$Xe and protons,
from comparison of only the measured \cite{Henzlova05} product
yields from reactions induced by $^{136}$Xe (Figs.\ 16 and 17)
with calculations by different versions of LAQGSM it
is difficult, if not impossible, to reveal the ``real" mechanisms
of nuclear reactions contributing to the production of measured isotopes.
\\

\vspace*{-10mm}

\begin{figure}[ht]                                                 
\centering
\hspace*{-5mm}
\includegraphics[height=210mm,angle=-0]{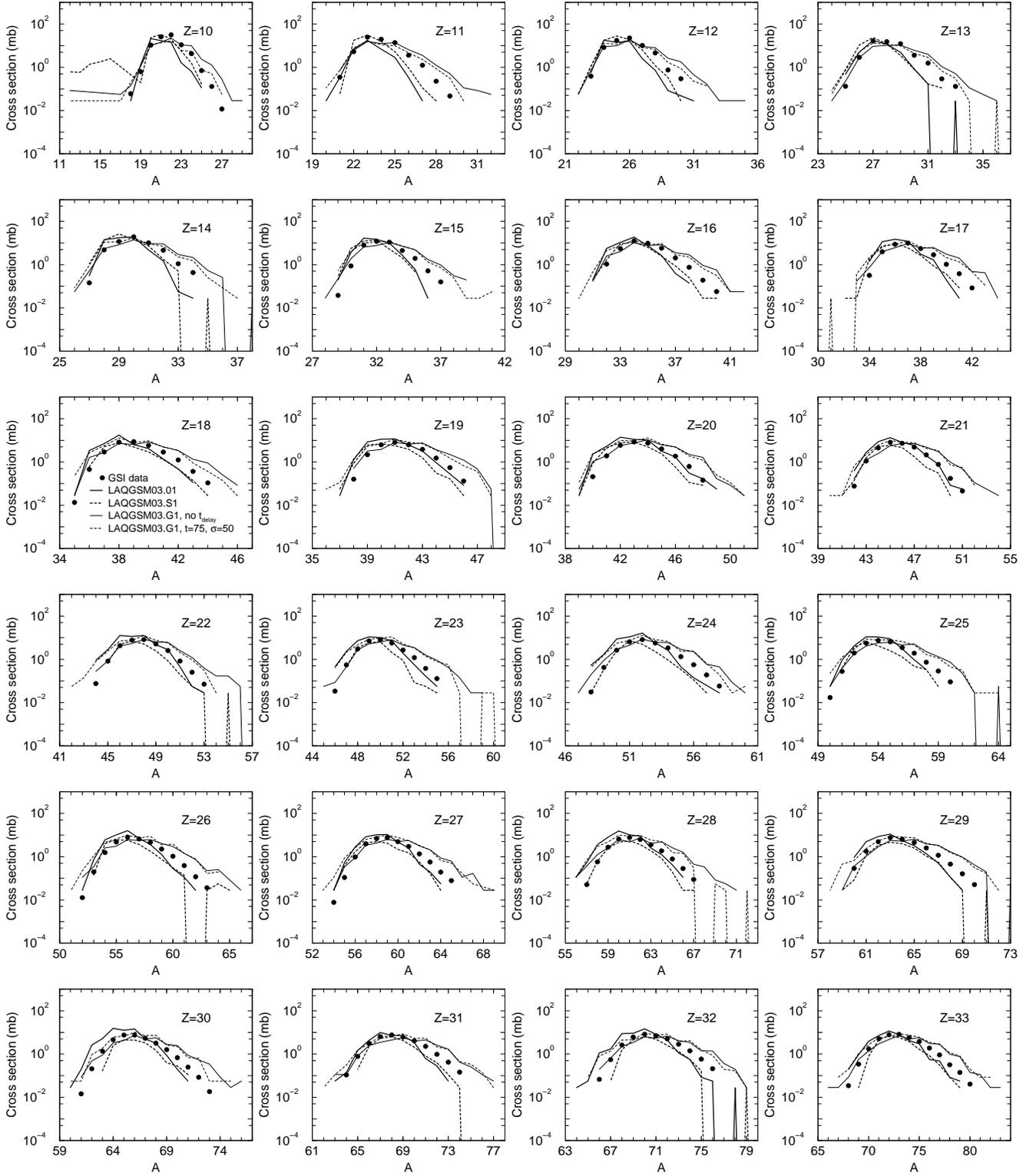}
\caption{
Comparison of LAQGSM03.01, LAQGSM03.S1, and LAQGSM03.G1
results (lines) for the total cross sections (yields) of nuclides
with Z from 10 to 33 produced from fragmentation of $^{136}$Xe
in 1 GeV/A $^{136}$Xe + $^{208}$Pb collisions with the GSI
data 
\cite{Henzlova05}
(circles),
as indicated. 
Results by LAQGSM03.G1 calculated without a delay time in GEMINI
and with $t_{delay} = 75$ and $\sigma_{delay} = 50$ are shown by
solid and dashed thin lines, respectively.}
\end{figure}

\begin{figure}[ht]                                                 
\centering
\hspace*{-5mm}
\includegraphics[height=210mm,angle=-0]{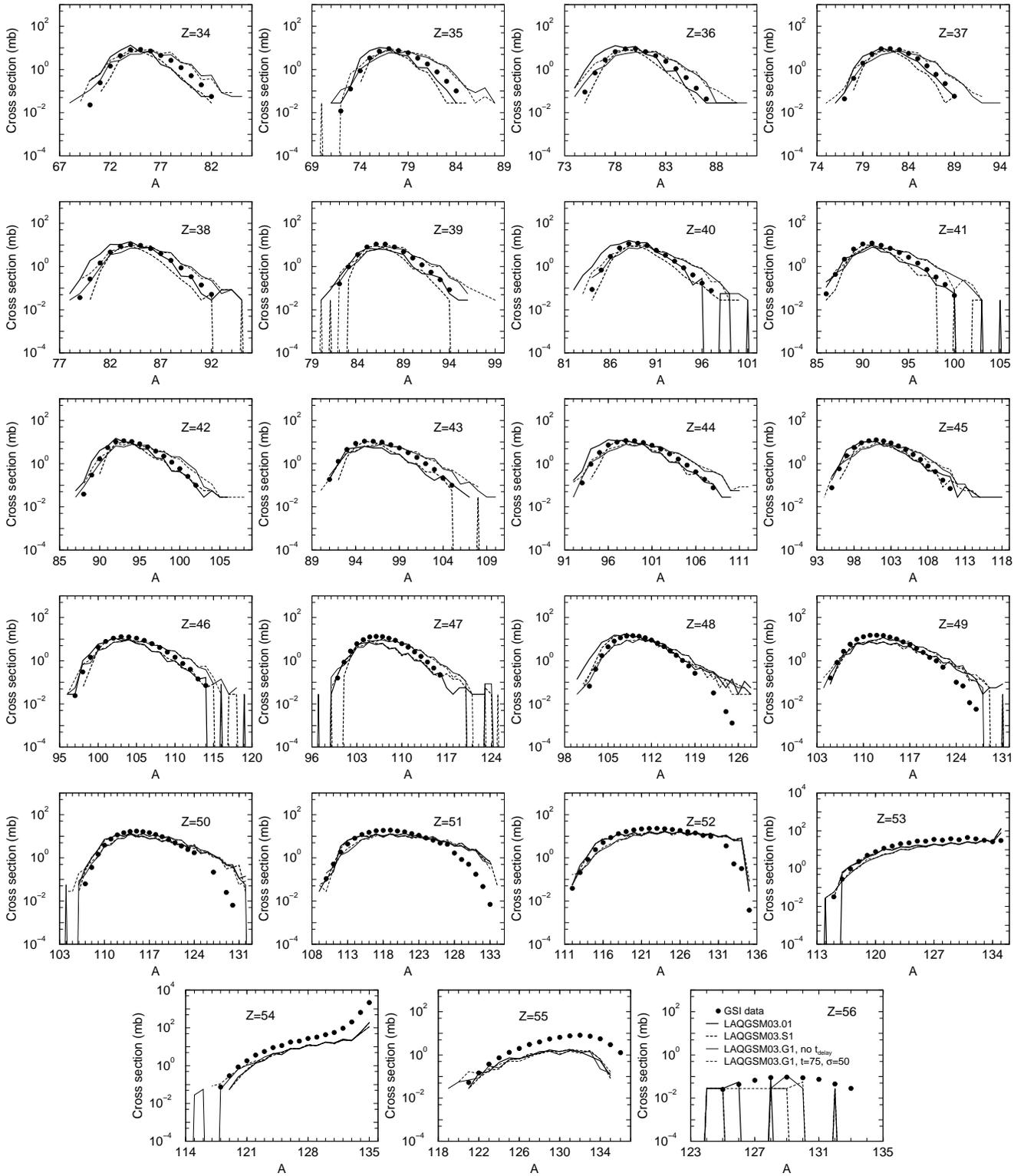}
\caption{The same as Fig.\ 16, but for products with Z from
34 to 56.}
\end{figure}
\clearpage            

{\noindent
This is why for reactions induced by  $^{136}$Xe,
we look additionally
at several kinematic characteristics like $<T_{kin}>$,
$<v_z>$, $<\Theta >$, and $<R>$, as we did
for proton-induced reactions (Figs.\ 1, 2, and  5).
}
The left panels of plots in Figs.\ 18 and 19 show such
characteristics 
(plus, the  $Z$-integrated mass product yield and the
cross section for the production of nuclides with $Z = 56$, 
shown in Fig.\ 18) calculated in the projectile frame of reference,
as all reactions at GSI are measured \cite{Henzlova05}. 
There is a big difference between
results of the standard version of LAQGSM and of its ``G" and
``S" versions for the calculated $Z$-integrated yield of isotopes
with $15 < A < 31$,  for $<T_{kin}>$ of isotopes with $20 < A < 80$,
and for $<v_z>$, $<\Theta >$, and especially for $<R>$ of almost
all products.
Unfortunately, none of these quantities were measured at GSI
so we can not identify a specific reaction mechanism
based on these results until experimental data are available.

To reveal the effect of angular momentum, $L$,
of the compound nucleus on the kinematic characteristics  $<v_z>$,
$<\Theta>$, and $<R>$
of the reaction induced by $^{136}$Xe, we have performed additional 
calculations with LAQGSM03.G1 assuming the angular momentum of all
compound nuclei is equal to zero. Results of such a modification
of LAQGSM03.G1 are shown in Fig.\ 19 with blue dotted thin lines, 
to be compared with
the thin black solid lines showing results from LAQGSM03.G1 considering 
the real angular momenta of all compound nuclei
(both of these calculations have no delay time in GEMINI).
The effect of angular 
momentum, $L$, of the compound nuclei on results for
$<\Theta>$, $<R>$, and $<v_z>$
calculated by GEMINI in LAQGSM03.G1 is not significant.

Together with results 
for products from only fragmentation of the projectile calculated
in the projectile frame of reference
(as all measurements at GSI are done)
shown on the left panels of Figs.\ 18 and 19, 
on the right panels of the same figures, we show side-by-side similar 
results calculated in the laboratory system that include isotopes
produced from both the projectile and the target. This is the way
a reaction really happens in nature, and how a transport code
``sees" and uses it in transport calculations from results
provided by event generators like LAQGSM.
Unfortunately, none of the currently available experimental
techniques allow the measurement of all products of heavy-ion reactions,
produced from both the projectiles and targets. We can not yet compare
the results shown on the right panels with any measurements.
However, we find them quite interesting and informative 
for nuclear applications, including to
users and developers of transport codes. 

Several phenomenological systematics are presently available in the
literature to estimate the cross sections of products from the fragmentation
of the projectile in a heavy-ion reaction, with the most advanced and
often used, especially at GSI, the EPAX parameterization by
K. S\"{u}mmerer and  B. Blank \cite{EPAX}. Such systematics are 
very fast and easy to calculate; they are 
useful and provide quite reliable results 
to estimate the fragmentation of the projectile in a
heavy-ion reaction, especially if experimental data for that
reaction, or for a not too different one, were used in deriving the
phenomenological systematics. But phenomenological systematics may
not provide reliable results for unmeasured reactions that differ
significantly from those used in fitting its parameters.
In addition, one should be very careful when using systematics
like EPAX in applications; the point is that most applications need
all products from a reaction, in the laboratory system, while EPAX
provides results from only fragmentation of the projectile.
So for the reaction  $^{136}$Xe + $^{208}$Pb discussed here, 
we can not use EPAX to calculate the inverse reaction $^{208}$Pb + $^{136}$Xe
adding the results with the ones obtained for the direct Xe+Pb reaction
with a hope to obtain all products from this reaction in the laboratory 
\\

\vspace*{-10mm}

\begin{figure}[ht]                                                 
\centering
\hspace*{-10mm}
\includegraphics[height=190mm,angle=-0]{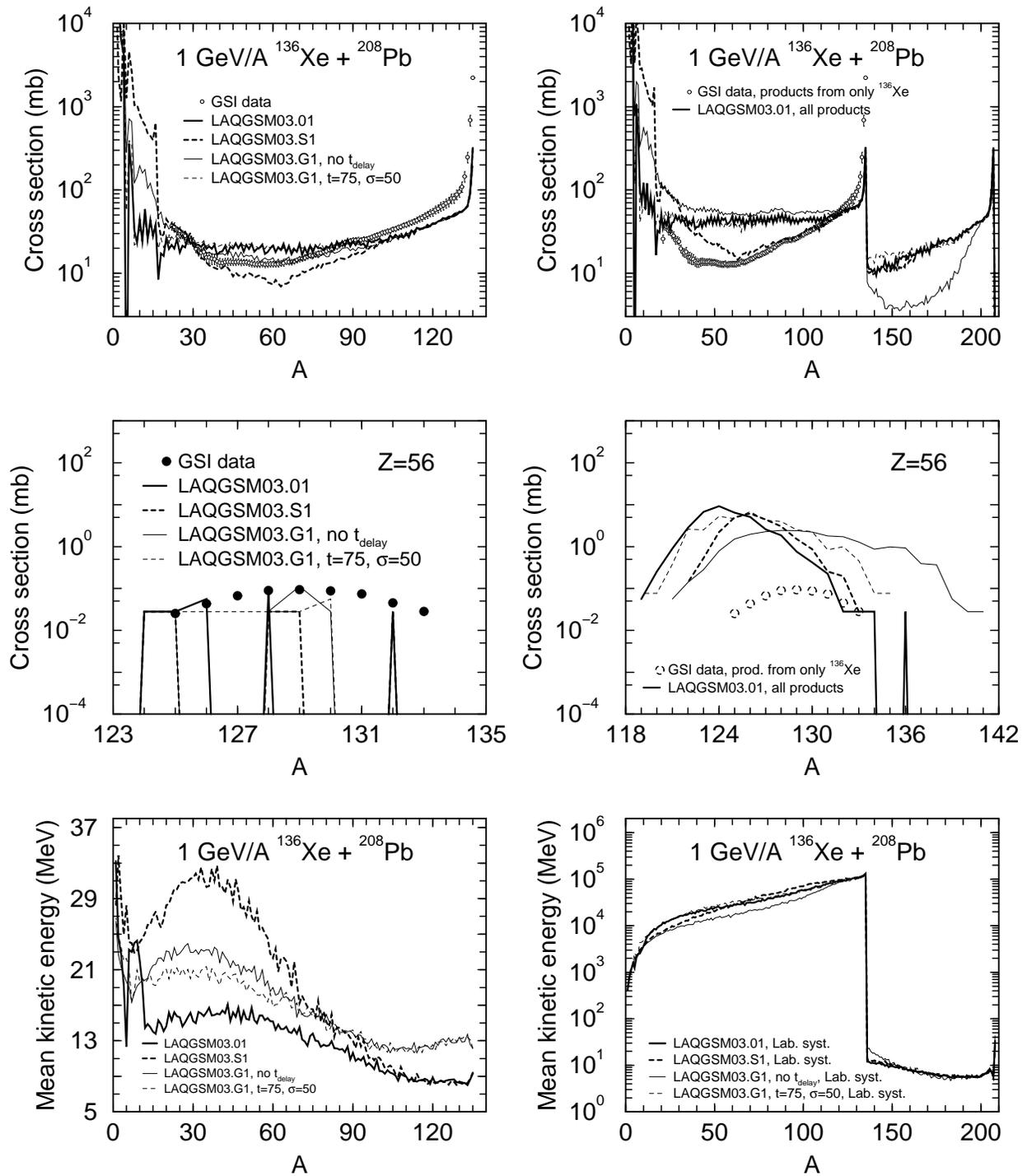}
\caption{{\bf Left panel:} Predictions of  LAQGSM03.01, LAQGSM03.S1, and 
LAQGSM03.G1
for the $Z$-integrated mass product yield,
cross section of the production of nuclides with $Z = 56$, and  
the mean kinetic energy of all products from the fragmentation of $^{136}$Xe
(in the beam system) from the 
1 GeV/A $^{136}$Xe + $^{208}$Pb reaction (lines)
compared with available experimental data (circles) 
\cite{Henzlova05},
as indicated.
{\bf Right panel:} The same as on the left panel, but calculated
in the laboratory system, as ``seen" by a transport code,
for all nuclides produced from both the
target, $^{136}$Xe, and the projectile,  $^{208}$Pb.
Experimental data (dashed circles on the right panel) are measured in the 
beam system and should be compared only with the results showed
on the left panel; this is why they do not agree with the laboratory
system results to be used by transport codes
shown in the right panel.
}
\end{figure}
\clearpage            

\begin{figure}[ht]                                                 
\centering
\hspace*{-10mm}
\includegraphics[height=190mm,angle=-0]{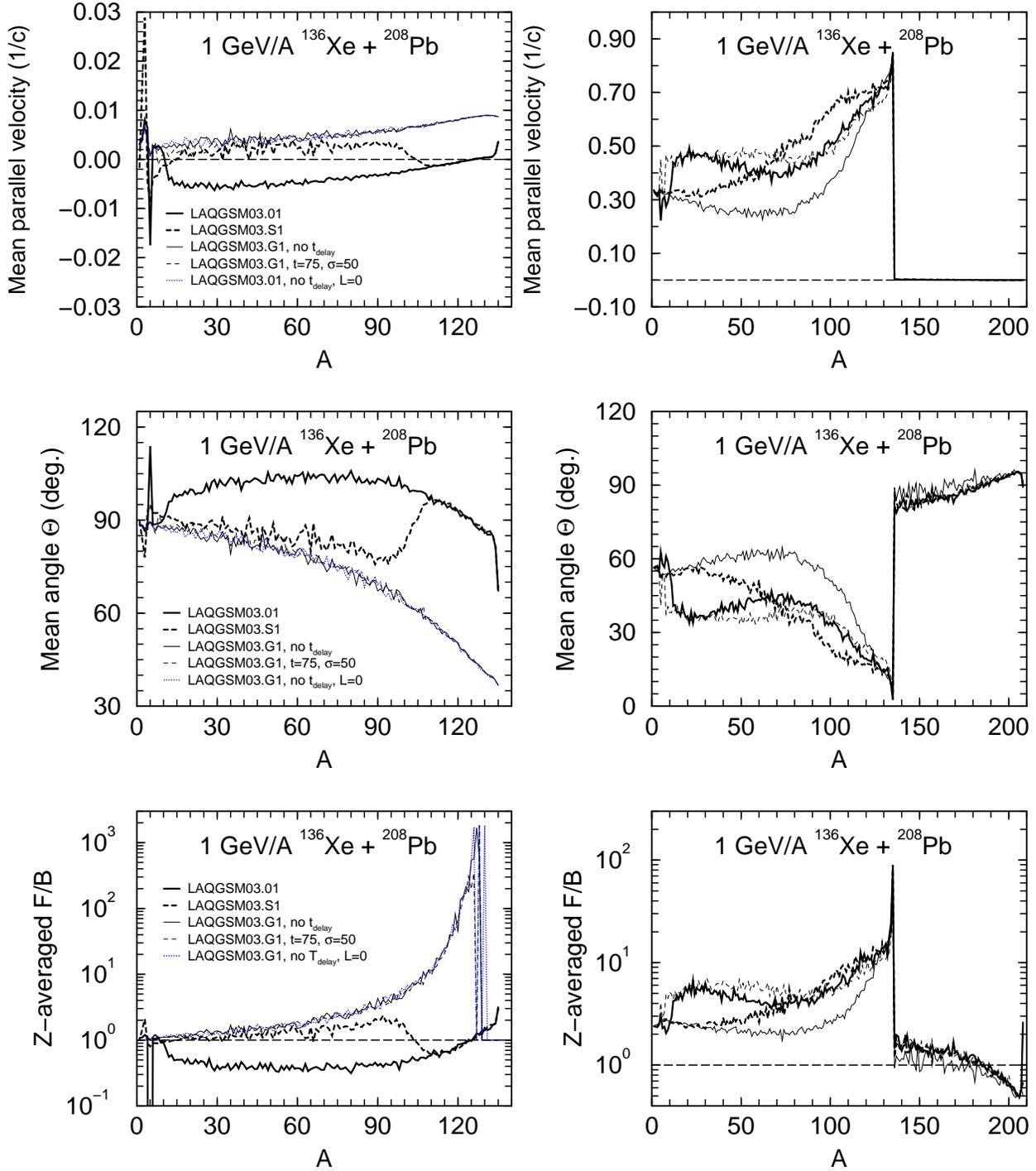}
\caption{The same as in Fig.\ 18, but for the
mean parallel velocity $v_z$, mean production angle $\Theta$,
Z-averaged A-dependence of the F/B ratio of the 
forward product cross sections to the backward ones.
To reveal the effect of angular momentum, $L$, of the
compound nucleus on results calculated by GEMINI in LAQGSM03.G1,
the blue dotted thin lines on the left panel
show results obtained assuming $L = 0$ in GEMINI, 
which should be compared with the results shown on the same plots
by thin solid black lines obtained using actual values of $L$.
}
\end{figure}
\clearpage            

{\noindent
system exactly as happens in reality and calculated by an event generator.
The point is that we have as products from a heavy-ion reaction not only 
those from the fragmentation of the projectile but also from the target.
}
Depending on the incident energy of the projectile and on the impact
parameter of the colliding nuclei, we may have a significant contribution
from intermediate systems
produced via the ``fusion" of a part of the projectile with a part of the
target. Such processes are missed by EPAX, while event generators
like our LAQGSM account for them. This is why we need reliable
event generators rather than phenomenological systematics in applications.
We hope to address this subject in more detail in a later paper.

\begin{center}
{\bf 3. Summary}
\end{center}
The recent 660 MeV p + $^{129}$I and 3.65 GeV p + $^{112}$Sn JINR  
activation measurements,
the new COSY H, He, Li, and Be production
data with  1.2 GeV protons on 13 nuclei from Al to Th,
the 300 MeV and 1 GeV p + $^{56}$Fe data measured at GSI in
inverse kinematics, and the new GSI data on fragmentation of Xe
from 1 GeV/nucleon $^{124}$Xe and $^{136}$Xe + Pb have
been analyzed with the standard versions 
(that use the Generalized Evaporation Model GEM2 of Furihata
to describe evaporation and fission)
of our event generators CEM03.01 and LAQGSM03.01,
as well as with their ``S" 
(which consider also multifragmentation of 
compound nuclei produced after 
the preequilibrium stage of reactions when their
excitation energies is above
$2\times A$ MeV) and ``G"
(which describe evaporation/fission stages of reactions using the
fission-like binary-decay model GEMINI of Charity {\it et al.}
instead of using GEM2)  modifications.
We conclude that from comparison of only these measured product
yields with calculations by different versions of our codes it
is difficult, if not impossible to uncover the ``real" mechanisms
of nuclear reactions contributing to the production of measured isotopes.
We find that some kinematic characteristics of nuclear
reactions like the the mean production angle $< \Theta >$, 
Z-averaged A-dependence of the F/B ratio of the 
forward product cross sections to the backward ones $< R >$,
and the mean kinetic energy of all products in the laboratory system,
$<T_{kin}>$ are described quite differently by GEM2, SMM, and GEMINI,
and may be a more powerful tool to understand  the ``real" mechasnisms
of fragment production. Such characteristics can not be measured
with the GSI inverse kinematics technique
or by the activation method used at JINR, but 
may be measured with some other techniques.
We encurage future measurements of such characteristcs both for
proton-induced and heavy-ion reactions.

We are grateful
to Prof.\ Robert Charity and Dr.\ Alexander Botvina for generously 
providing us their GEMINI and SMM codes implemented into the
``G" and ``S" versions of our event generators.
We thank our collaborators, Drs.\ Arnold Sierk, Richard Prael,
and Nikolai Mokhov for their important contributions and
support of our work on development the ``S" and ``G" versions
of our codes, as well as for useful discussions of the results.
This work was supported by the 
Advanced Simulating Computing (ASC) Program at 
the Los Alamos National Laboratory 
operated by the University of California for the
US Department of Energy.

\end{document}